\documentclass[prb,aps,amsmath,amssymb,amsfonts,superscriptaddress,showpacs,twocolumn]{revtex4}
\usepackage{amsmath}
\usepackage{graphicx}
\usepackage{epsfig}
\def\moy#1{\left\langle #1 \right\rangle}

\def\beq{\begin{equation}}
\def\eeq{\end{equation}}
\def\be{\begin{equation}}
\def\ee{\end{equation}}

\def\cG0{{\cal G}_0}
\def\cG{{\cal G}}

\def\vacu{|{\rm vac}\rangle}
\def\zk{Z({\mathbf{k}})}

\def\spinup{\uparrow}
\def\spindown{\downarrow}
\def\vk{\mathbf{k}}
\def\vK{\mathbf{K}}

\def\vR{{\bf R}}

\def\matZ{\hat{Z}}
\def\matSigma{\hat{\Sigma}}
\def\efat{\mbox{\boldmath$\varepsilon$}}
\def\a{\alpha}
\def\b{\beta}
\def\d{\delta}

\def\eo{\varepsilon^0}
\def\g{\gamma}
\def\G{\Gamma}
\def\l{\lambda}

\def\o{\omega}

\def\s{\sigma}

\def\hn{\hat{n}}

\def\tp{t^\prime}
\def\uc2{$U_{c2}$}
\def\uc1{$U_{c1}$}

\def\dd{d^{\,\dagger}}
\def\ud{\underline{d}}
\def\udd{\underline{d}^\dagger}
\def\fd{f^\dagger}
\def\ftd{\widetilde{f}^{\,\dagger}}

\def\Hloc{H_{\rm{loc}}}
\def\DD{\hat{\Delta}}
\def\bra{\langle}
\def\ket{\rangle}
\def\braA{\langle A|}

\def\bran{\langle n|}

\def\ketA{|A\rangle}
\def\kettA{|\widetilde{A}\rangle}
\def\ketB{|B\rangle}
\def\ketC{|C\rangle}

\def\ketn{|n\rangle}
\def\ketm{|m\rangle}
\def\ketal{|\alpha\rangle}
\def\brabe{\langle\beta|}
\def\phisb{\phi^{\hfill}}
\def\phidn{\phi^\dagger_n}
\def\phin{\phi_n^{\hfill}}
\def\phim{\phi_m^{\hfill}}

\def\phiAn{\phi_{An}^{\hfill}}

\def\phiBn{\phi_{Bn}^{\hfill}}

\def\phiGn{\phi_{\,\Gamma n}^{\hfill}}
\def\phiG{\phi_{\,\Gamma}}
\def\GroupeEquations#1{\begin{subequations}  #1  \end{subequations}}
\def\moy#1{\left\langle #1 \right\rangle}

\newcommand{\empile}[2]{\genfrac{}{}{0pt}{}{#1}{#2}}
\begin{document}
\title{Rotationally-invariant slave-boson formalism\\
and momentum dependence of the quasiparticle weight}
\author{Frank Lechermann}
\email{Frank.Lechermann@physnet.uni-hamburg.de}
\affiliation{I. Institut f\"{u}r Theoretische Physik,
Universit\"{a}t Hamburg,Jungiusstrasse 9,20355 Hamburg, Germany}
\affiliation{Centre de Physique Th\'eorique, \'Ecole Polytechnique,
91128 Palaiseau Cedex, France}
\author{Antoine Georges}
\affiliation{Centre de Physique Th\'eorique, \'Ecole Polytechnique,
91128 Palaiseau Cedex, France}
\author{Gabriel Kotliar}
\affiliation{Serin Physics Laboratories, Rutgers University, Piscataway, NJ, USA}
\affiliation{Centre de Physique Th\'eorique, \'Ecole Polytechnique,
91128 Palaiseau Cedex, France}
\author{Olivier Parcollet}
\affiliation{Service de Physique Th\'eorique, CEA/DSM/SPhT-CNRS/SPM/URA 2306
CEA Saclay, F-91191 Gif-Sur-Yvette, France}
\begin{abstract}
We generalize the rotationally-invariant formulation of the slave-boson
formalism to multiorbital models, with arbitrary interactions, crystal fields,
and multiplet structure. This allows for the study of multiplet effects
on the nature of low-energy quasiparticles. Non-diagonal components of the matrix 
of quasiparticle weights can be calculated within this framework. When combined 
with cluster extensions of dynamical mean-field theory, this method
allows us to address the effects of spatial correlations, such as the generation 
of the superexchange and the momentum dependence of the quasiparticle weight.
We illustrate the method on a two-band Hubbard model, a Hubbard model made of two 
coupled layers, and a two-dimensional single-band Hubbard model (within
a two-site cellular dynamical mean-field approximation).
\end{abstract}
\maketitle


\section{Introduction and Motivations}
\label{sec:intro}

\subsection{General motivations}
\label{sec:motivations}

The method of introducing auxiliary bosons in order to facilitate the description 
of interacting fermionic systems is an important technique in theoretical 
many-body physics. In this regard,  the so-called slave boson (SB) approach is a 
very useful tool in dealing with models of 
strongly correlated electrons. Slave boson mean-field theory (SBMFT), i.e., at the 
saddle-point level, is the simplest possible realization of a Landau Fermi liquid
(for a review, see e.g.~Ref.\onlinecite{kotliar_largeN_leshouches_1995}). Within 
SBMFT, a simplified description of the low-energy quasiparticles is obtained, 
while high-energy (incoherent) excitations are associated with fluctuations around
the saddle point. In particular, two essential features are captured by SBMFT:
(i) the Fermi surface (FS) of the interacting system (satisfying Luttinger's 
theorem) is determined by the zero-frequency self-energy, which is in turn 
determined by the Lagrange multipliers associated with the constraints and (ii) 
the quasiparticle (QP) weight $Z$ is determined by the saddle-point values of the 
slave bosons. Hence SBMFT is a well-tailored technique when attempting to 
understand the low-energy physics emerging from more sophisticated theoretical 
tools, such as for example dynamical mean-field theory (DMFT), which deals with 
the full frequency dependence of the self-energy.

In this paper, we are concerned with the construction of a slave-boson
formalism which is able to deal with the two following problems.

1) In multi-orbital models, handle an arbitrary form of the interaction
hamiltonian, not restricted to density-density terms, and possibly including
interorbital hoppings or hybridizations. We aim in particular at describing
correctly the multiplets (eigenstates of the atomic hamiltonian), but we also want
to be able to work in an arbitrary basis set, not necessarily that of the atomic
multiplets (and of course, to obtain identical results, independent of the choice 
of basis).

2) Describe situations in which the QP weight is 
{\it not uniform along the Fermi surface}, but instead varies as a function 
of the momentum, i.e., $Z$=$\zk$.

There are clear physical motivations for addressing each of these issues.
The first one is encountered whenever one wants to deal with a specific 
correlated material in a realistic setting 
(see e.g. Ref.~\onlinecite{imada_mit_review}). Usually, more than one band is 
relevant to the physics (e.g. a $t_{2g}$ triplet or $e_g$ doublet for transition 
metal-oxides, or the full $7$-fold set of $f$-orbitals in rare-earth, actinides
and their compounds). The second issue is an outstanding one in connection with 
cuprate superconductors. In those materials, a strong differentiation in 
momentum space is observed in the ``normal'' (i.e., non-superconducting) state, 
especially in the underdoped regime (for a review, see e.g. 
Ref.~\onlinecite{damascelli_rmp_2003}). For momenta close to the nodal regions, 
i.e., close to the regions where the superconducting gap vanishes, reasonably 
long-lived QPs are found. In contrast, in the antinodal directions, 
the angle-resolved photoemission (ARPES) spectra reveal only a broad lineshape 
with no well-defined QPs. The
nature of the incipient normal state in the underdoped regime (i.e., the
state achieved by suppressing the intervening superconductivity) has been
a subject of debate. One possibility is that QPs would
eventually emerge at low-enough temperature in the antinodal region as
well, but with a much smaller QP weight $Z_{\rm{AN}}$$\ll$$Z_{\rm{N}}$. 
Another possibility is that coherent QPs simply do
not emerge in the antinodal region. Anyhow, there is evidence from ARPES
and other experiments~\cite{damascelli_rmp_2003} that the QP
weight (whenever it can be defined) has significant variation along the FS
and is larger at the nodes. Since the QP weight sets the scale for the coherence 
temperature below which long-lived QPs form, a smaller $Z$ means a smaller 
coherence temperature. Hence, if the temperature is higher than the coherence 
scale associated with momenta close to the nodes, and larger than the one 
associated with the antinodes, QPs will be visible only in the nodal regions. 
At this temperature, the FS will thus appear as being formed of ``Fermi arcs'', 
as indeed observed 
experimentally~\cite{norman_arcs_nature_1998}. Important differences between the 
nodal and antinodal region in the superconducting state have also been unraveled 
by recent experiments, in particular from Raman scattering which revealed two 
different energy scales with different doping dependence, associated
with each of these regions~\cite{letacon_raman_natphys}.
Momentum-space differentiation of QP properties is therefore a key feature of 
cuprate superconductors, but it is also an issue which is particularly difficult 
to handle theoretically.

As we now explain, these two issues are actually closely related one to the other.
In a general multi-orbital model, the self-energy is a matrix
$\Sigma_{\a\b}$ ($\a,\b$ are orbital indices). Except when a particular symmetry
dictates otherwise, this matrix has in general off-diagonal (interorbital)
components and these off-diagonal components may have a non-zero linear term in
the low-frequency expansion, hence yielding non-diagonal components of the
matrix of QP weights defined as:
\begin{equation}
\matZ=\left[1-\frac{\partial}{\partial\omega}\matSigma\,\right]_{\o=0}^{-1}
\quad.\label{eq:defZ}
\end{equation}
On the other hand, a momentum dependent QP weight $\zk$ means that, in real-space,
$Z_{ij}$$=$$Z(\vR_i-\vR_j)$ depends on the separation between lattice sites 
(a momentum-independent $Z$ means that $Z_{ij}$$=$$Z\,\delta_{ij}$ is purely 
local).
Hence, in both cases, one has to handle a QP weight which is a matrix in either the
orbital or the site indices. The connection becomes very direct in the
framework of cluster extensions of DMFT
(for reviews, see e.g. 
Refs~[\onlinecite{georges_review_dmft,maier_cluster_rmp_2005,
kotliar_review_rmp_2006,tremblay_review_jltp_2006}]). There, a lattice
problem is mapped onto a finite-size cluster which is self-consistently
coupled to an environment. This finite-size cluster can be viewed as a 
multi-orbital (or molecular) quantum impurity problem, in which each site plays 
the role of an atomic orbital.
Recently, numerical solutions of various forms of cluster extensions to the
DMFT equations for the two-dimensional Hubbard model have clearly revealed the
phenomenon of momentum-space
differentiation~\cite{senechal_hotspots_prl_2004,par04,civelli_breakup_prl_2005}.
Developing low-energy analytical
tools to interpret, understand, and generalize the results of these calculations
is clearly an important and timely issue. The slave-boson methods developed
in the present work are a step in this direction.

Obviously, the existence of off-diagonal components of the $\hat{Z}$-matrix is
a basis-set dependent issue. A proper choice of orbital basis can be made,
which diagonalizes this matrix. In certain cases, this basis is dictated
by symmetry considerations, while in the absence of symmetries,
the basis set in which $Z$ is diagonal cannot be guessed a priori.
For instance, in a two-site cluster or two-orbital model
in which the two sites play equivalent roles, even and odd combinations diagonalize
not only the $\hat{Z}$-matrix, but in fact the self-energy matrix
itself for all frequencies (see Sec.~\ref{sec:results}). In such cases, it may be 
favorable to work in this orbital basis set, and deal only with diagonal QP 
weights. However, performing the rotation into this orbital basis set will in 
general transform the interacting hamiltonian into a more complicated form. For 
example, starting from a density-density interaction, it may induce interaction 
terms which are not of the density-density type (i.e., involve exchange, pair 
hopping, etc.). For these reasons, it is essential to consider slave-boson 
formalisms which can handle both arbitrary interaction terms, and non-diagonal 
components of the QP weight matrix: these two issues are indeed connected.
The formalism presented in this article builds on earlier ideas
of Li, W\"{o}lfle, Hirschfeld
and Fr\'esard~\cite{li_rotinv1_prb_1989,fresard_rotinv_intjmodphys_1992}
(see Appendix.~\ref{appx:single_orbital}),
in which the SB formalism is formulated in a 
{\it fully rotationally-invariant manner}
(see also Refs~[\onlinecite{attaccalite_gutzwiller_prb_2003,ferrero_thesis_2006}] 
in the framework of the Gutzwiller approximation),
so that the orbital basis set needs not be specified from the beginning, and the
final results are guaranteed to be equivalent irrespectively of the chosen
basis set.

\subsection{Some notations}
\label{sec:notations}

In this paper, we shall consider multi-orbital models of correlated
electrons with hamiltonians of the form:
\begin{eqnarray}
H&=&H_{\rm{kin}}\,+\,\sum_i\,\Hloc[i]\label{eq:ham_generic}\quad,\\
\mbox{with}&&\quad
H_{\rm{kin}}=\sum_\vk \sum_{\a\b} \varepsilon_{\a\b}(\vk)\,
\dd_{\vk\a} d_{\vk\b}^{\hfill}\quad.\label{eq:ham_kin}
\end{eqnarray}
In these expressions, $\a,\b$ label electronic species and run from $1$ to $M$
(i.e., $M$ is twice the number of atomic orbitals in the context of a multi-orbital
model of electrons with spin: $\a$$=$$(m,\s),$$\s$$=$$\spinup,\spindown$).
The $k$ vector runs over the Brillouin zone of the lattice, whose sites are 
labeled by $i$ (in the context of cluster-DMFT, $i$ will label clusters and runs 
over the superlattice sites, thus $\vk$ runs over the reduced Brillouin zone of 
the superlattice, see Sec.~\ref{sec:cdmft}).
The first term in (\ref{eq:ham_generic}) is the kinetic energy: 
$\varepsilon_{\a\b}(\vk)$ is the Fourier transform of the
(possibly off-diagonal) hoppings and does not contain any local terms
(i.e., $\sum_\vk\varepsilon_{\a\b}(\vk)$$=$0). $\Hloc$ contains both the one-body 
local terms and the interactions, assumed to be local. A general form for $\Hloc$
is~\footnote{In this paper, the chemical potential is usually included
in the one-body part $\varepsilon^0_{\a\b}$ of the local hamiltonian.}:
\begin{equation}
\Hloc = \sum_{\a\b} \varepsilon^0_{\a\b} \dd_\a d_\b^{\hfill} +\frac{1}{2}
\sum_{\a\b\g\d}\,U_{\a\b\g\d}\,\dd_\a\dd_\b d_\d^{\hfill} d_\g^{\hfill}\quad.
\label{eq:hloc_generic}
\end{equation}
Fock states form a convenient basis set of the local Hilbert space on each site.
They are specified by sequences $n$$=$$(n_1,\ldots,n_M)$, with $n_\a$$=$$0,1$
(we consider a single site and drop the site index):
\beq
|n\ket = \left(\dd_1\right)^{n_1} \cdots \left(\dd_M\right)^{n_M} \vacu\quad.
\label{eq:def_fockstates}
\eeq
In the following, $\{|A\ket\}$ will denote an arbitrary basis set of the local
Hilbert space, specified by its components on the Fock states:
\beq
|A\ket = \sum_n \bra n|A\ket\,|n\ket\quad,
\eeq
while $|\G\ket$ will denote the eigenstates of the local hamiltonian, i.e., the
`atomic' multiplets such that:
\beq
\Hloc|\G\ket\,=\,E_\G\,|\G\ket\quad.
\eeq

\subsection{Slave bosons for density-density interactions: a reminder}
\label{sec:KR_bosons}

When the orbital densities $n_\a$ are good quantum numbers for the
local hamiltonian $\Hloc$, i.e., when the eigenstates of the latter are
labeled by $n_\a$, a very simple slave boson formalism can be constructed which
is a direct multi-orbital generalization of the 4-boson scheme
introduced by Kotliar and 
Ruckenstein~\cite{kotliar_4bosons_prl_1986,fresard_multiorbital_prb_1997}.
While this is standard material, we feel appropriate to briefly remind the reader 
of how this scheme works, in order to consider generalizations later on. We thus 
specialize in this subsection to a local hamiltonian of the form:
\begin{equation}
\Hloc = \sum_{\a} \varepsilon^0_{\a} \hn_\a \,+\,
\sum_{\a\b}\,U_{\a\b}\,\hn_\a\,\hn_\b\quad,
\label{eq:hloc_density}
\end{equation}
so that the multiplets are the Fock states $|n\ket$ themselves, with 
eigenenergies:
\beq
E_n = \sum_{\a} \varepsilon^0_{\a} n_\a \,+\,
\sum_{\a\b}\,U_{\a\b}\,n_\a\,n_\b\quad.
\label{eq:energ_atomic_simple}
\eeq

To each Fock state, one associates a boson creation operator $\phi^\dagger_n$.
Furthermore, auxiliary fermions $\fd_\a$ are introduced which correspond to
{\it quasiparticle degrees of freedom}. The (local) enlarged Hilbert space 
thus consists of states which are built from tensor products of a QP Fock state, 
times an arbitrary number of bosons. In contrast, the 
{\it physical Hilbert space} is generated by the basis set consisting of the
$2^M$ states which contain exactly one boson, and in which this boson matches
the QP Fock state. Thus, the states representing the original physical states
(\ref{eq:def_fockstates}) in the enlarged Hilbert space, in a one-to-one manner, 
are the following (``physical'') states:
\beq
|\underline{n}\,\ket\,\equiv\,\phidn\vacu \otimes |n\ket_f\quad.
\label{eq:states_simple}
\eeq
The underlining in $|\underline{n}\,\ket$ allows one to distinguish between the
original Fock state of the physical electrons $|n\ket$, and its representative 
state in the enlarged Hilbert space.
In this expression, $|n\ket_f$ stands for the QP Fock state:
\beq
|n\ket_f \equiv \left(\fd_1\right)^{n_1} \cdots \left(\fd_M\right)^{n_M}\vacu\quad.
\label{eq:def_FockQP}
\eeq

It is easily checked that a simple set of constraints uniquely specifies the 
physical states among all the states of the enlarged Hilbert space, namely:
\begin{eqnarray}
\sum_n \phidn\phin&=& 1
\label{eq:constraint1_simple}\\
\sum_n n_\a\, \phidn\phin&=& \fd_\a f_\a^{\hfill} 
\quad,\qquad\forall\,\a\quad.
\label{eq:constraint2_simple}
\end{eqnarray}
The first constraint imposes that only states with a single boson are retained,
while the second one insures that the fermionic (QP) and bosonic contents match.
Obviously, the saddle-point values of the slave bosons will have a simple 
interpretation, $|\phin|^2$ being the probability associated with the Fock space 
configuration $n$.

The operator:
\begin{equation}
\underline{d}^\dagger_{\,\alpha} = \sum_{nm} \bra n|\fd_\a|m\ket\, 
\phi^\dagger_n\phi_m\, \fd_\a
\label{eq:creation_op_KR}
\end{equation}
is a faithful representation of the physical electron creation operator on the
representatives (\ref{eq:states_simple}), namely:
\beq
\udd_{\a} |\underline{n}\,\ket = \sum_{n^\prime} \bra n^\prime|\dd_a|n\ket\,
|\underline{n}^{\prime}\,\ket\quad,
\eeq
in which, in fact, the r.h.s is either zero (if $n_\a$$=1$) or composed of just a
single state (with $n^\prime_\a$$=1$ and otherwise $n_\b$$=$$n^\prime_\b$ for
$\b$$\neq$$\a$). This expression of the physical electron operators is not unique
however: obviously, one can for example multiply this with any operator acting as 
the identity on the physical states. This is true as long as the constraint is 
treated exactly. When treated in the mean-field approximation however,
(i.e., at saddle point), these equivalent expressions will not lead to the same 
results. In fact, (\ref{eq:creation_op_KR}) suffers from a serious drawback namely
it does not yield the exact non-interacting ($U_{\a\b}$$=$$0$) limit at 
saddle-point. Instead, the expression:
\begin{eqnarray}
&&\hspace{-0.5cm}\udd_{\,\alpha} = \sum_{nm} \bra n|\fd_\a|m\ket\,
[\DD_\a]^{-\frac{1}{2}}\phi^\dagger_n \phim\,[1-\DD_\a]^{-\frac{1}{2}}\,
\fd_\a \label{eq:creation_op_KR_norm}\\
&&\mbox{with}\quad\DD_\a[\phi]\equiv\sum_{n} n_\a\,\phi^\dagger_{n}\phin\quad,
\end{eqnarray}
turns out to satisfy this requirement, while having exactly the
same action as (\ref{eq:creation_op_KR}) when acting on physical states.
This choice of normalization is actually very natural given the
probabilistic interpretation of $|\phin|^2$: the expression
$[\DD_\a]^{-1/2}\phi^\dagger_n$ (resp. $\phim\,[1-\DD_\a]^{-1/2}$) is
actually a probability amplitude, normalized over the {\it restricted
set} of physical states such that $n_\a$$=$$1$ (resp. $n_\a$$=$$0$). Hence
the combination of boson fields in (\ref{eq:creation_op_KR_norm}) is
a transition probability between the state $m$ with $m_\a$$=$$0$ and
the state $n$ with $n_\a$$=$$1$.

Anyhow, whether the simplest expression (\ref{eq:creation_op_KR}) or the
normalized expression (\ref{eq:creation_op_KR_norm}) is chosen for the
physical operator, the relation between the physical and QP single-particle 
operators is of the form:
\beq
d_\a = \hat{r}_\a[\phi]\,f_\a
\label{eq:def_r}
\eeq
It is important to note that the orbital index carried by the physical
operator is identical to that of the QP operator. An immediate consequence is
that the self-energy at the saddle-point level is a {\it diagonal matrix in
orbital space} $\Sigma_{\a\b}$$=$$\delta_{\a\b}\Sigma_\a$, which reads:
\begin{eqnarray}
\Sigma_{\a}(\omega)&=&\Sigma_\a(0)+\omega\,\left(1-\frac{1}{Z_\a}\right)\quad,
\label{eq:Sigma_simple}\\
\mbox{with}\qquad\quad\quad
Z_\a &=& |r_\a|^2 \label{eq:Z_simple}\\
\Sigma_{\a}(0) &=& \lambda_\a/|r_\a|^2-\eo_\a\quad.
\label{eq:Sigma0_simple}
\end{eqnarray}
In these expressions, $r_\a$ is evaluated at saddle-point level, and
$\lambda_\a$ is the saddle-point value of the Lagrange multipliers enforcing
the constraint (\ref{eq:constraint2_simple}).

The expression (\ref{eq:Z_simple}) of the QP weight is an immediate consequence of
(\ref{eq:def_r}): at saddle-point level, $r_\a$ becomes a c-number and
(\ref{eq:def_r}) implies that the physical electron carries a spectral weight
$|r_\a|^2$. Hence, in order to describe within SBMFT situations in which the QP 
weight is a non-diagonal matrix, one must disentangle the orbital indices carried 
by the physical electron and those carried by the QP degrees of freedom.
These operators will then be related by a non-diagonal matrix:
\beq
d_\a\,=\,\hat{R}_{\a\b}[\phi]\,f_\b\quad.
\label{eq:defR}
\eeq
This is precisely what the formalism exposed in this article achieves.
The physical significance of such a non-diagonal relation is that creating a
physical electron in a given orbital may induce the creation of QPs
in any other orbital. Thinking of orbital as real-space indices (within e.g.
cluster-DMFT), this means that the creation of a physical electron on a given site
induces QPs on other sites in a non-local manner, corresponding to
a momentum-dependent $\zk$.

\subsection{Difficulties with naive generalizations to the multi-orbital case}
\label{sec:oops!}

Let us come back to the general multi-orbital interaction (\ref{eq:hloc_generic}).
In order to motivate the fully rotationally-invariant formalism exposed
in the next section, let us point out some difficulties arising when attempting 
to generalize the simple SB formalism of the previous section.

The central difference between the general interaction (\ref{eq:hloc_generic}) and
the density-density form (\ref{eq:hloc_density}) is that the atomic multiplets
$|\Gamma\ket$ are no longer Fock states. Thus, it would seem natural to associate
a slave boson $\phiG$ to each of the atomic multiplets. Indeed, 
B\"{u}nemann {\sl et al.}~\cite{bunemann_gutzwiller_prb_1998}
(see also [\onlinecite{attaccalite_gutzwiller_prb_2003}]) have proposed 
generalized Gutzwiller wave functions in which a variational parameter (a.k.a a 
probability $|\phiG^{\hfill}|^2$) is associated with each atomic
multiplet (see also Ref.~[\onlinecite{tre95}] and the recent work of
Dai {\sl et al.}~\cite{dai06} in the SB context).
A slave-boson formulation requires a clear identification of the physical
states within the enlarged Hilbert space. A natural idea is to define those
in one-to-one correspondence with the atomic multiplets, as:
\beq
|\underline{\Gamma}\ket\,\stackrel{?}{=}\,\phiG^\dagger\vacu\,\otimes\,
\sum_n \bra n|\G\ket\,|n\ket_f\quad.
\label{eq:states_naive_1}
\eeq
The local part of the hamiltonian has a simple representation on these physical 
states $\Hloc$$=$$\sum_\G E_\G \phiG^\dagger\phiG$. However, a major difficulty is
that there is no simple constraint implementing the restriction to these physical 
states, and such that it is quadratic in the fermionic (QP) degrees of freedom 
(which is essential in order to yield a manageable saddle point). In particular, 
it is easily checked that the apparently natural constraint~\cite{dai06}:
\beq
\fd_\a f_\a^{\hfill}\,\stackrel{?}{=}
\,\sum_\G\, \bra\G|\hat{n}_\a|\G\ket\,\phiG^\dagger\phiG
\label{eq:constraint_naive_1}
\eeq
is actually not satisfied by the states (\ref{eq:states_naive_1}) as an operator
identity~\footnote{Note however that both sides of (\ref{eq:constraint_naive_1})
have identical matrix elements between physical states}.
Further difficulties also arise when attempting to derive an expression for
the physical creation operators. These difficulties stem from the fact that
two atomic multiplets having particle numbers differing by one unit
cannot in general be related by the action of a single-fermion creation.

One might also think of defining the physical states in correspondence to the
Fock states, as:
\beq
|\underline{n}\ket\,\stackrel{?}{=}\,|n\ket_f \otimes
\sum_\G\,\bra\G|n\ket\,\phiG^\dagger\vacu
\label{eq:states_naive_2}
\eeq
which do satisfy the following quadratic constraint:
\beq
\fd_\a f_\a^{\hfill}\,=\,\sum_{\G\G^\prime}\,
\bra\G|\hat{n}_\a|\G^\prime\ket\,
\phiG^\dagger\phi_{\,\Gamma^\prime}^{\hfill}\quad.
\label{eq:constraint_naive_2}
\eeq
However, another difficulty then arises. Namely, it is not possible to write the
local interaction hamiltonian purely in terms of bosonic degrees of
freedom, which is the whole purpose of SB representations. In particular, the
obvious expression $\Hloc$$=$$\sum_\G E_\G \phiG^\dagger\phiG^{\hfill}$ which had 
the correct action on states (\ref{eq:states_naive_1}) no longer works for states
(\ref{eq:states_naive_2}) since it leaves unchanged the fermionic content of them.

After some thinking, one actually realizes that these naive generalizations are 
all faced with the same problem, namely that they do not embody the crucial 
conceptual distinction between physical and QP degrees of freedom. Both
(\ref{eq:states_naive_1}) and (\ref{eq:states_naive_2}) assume a
priori a definite relation between the physical and QP content
of a state. The key to a successful SB formalism is therefore to
{\it disentangle physical and quasiparticle degrees of freedom},
and letting the variational principle at saddle point decide which
relationship actually exists between those.

We shall see however in Sec.~\ref{sec:two_orbitals} that,
provided the local hamiltonian {\it has enough symmetries}, the 
rotationally-invariant formalism of the present article does correspond to 
assigning at saddle point a probability to each atomic configuration (multiplet)
$|\G\ket$, hence establishing contact with the previous works of
Refs.~[\onlinecite{bunemann_gutzwiller_prb_1998,dai06}]. Yet
for less symmetric hamiltonians, the general formalism of the
present article is requested.

\section{Rotationally-invariant slave-boson formalism}
\label{sec:formalism}

\subsection{Physical Hilbert space and constraints}
\label{sec:hilbert_spaces}

In order to construct a SB formalism in which physical and QP
states are disentangled, we shall associate a slave boson $\phiGn$
to {\it each pair} of atomic multiplet $|\G\ket$ and QP
Fock state $|n\ket_f$. More generally, we can work in an
arbitrary basis set $|A\ket$ of the local Hilbert space, not necessarily
that of the atomic multiplets, and consider slave bosons
$\phiAn$. As we shall see, the formalism introduced in this article is such
that two different choices of basis sets are related by a unitary transformation
and therefore lead to identical results. In particular,
one could also choose the physical Fock states $|m\ket_d$ as the basis
set $A$, and work with slave bosons $\phi_{mn}^{\hfill}$ which form the components
of a {\it density matrix} connecting the physical and QP spaces.
It is crucial however to keep in mind that the first index ($A$) refers to
{\it physical-electron states}, while the second one ($n$) refers
to {\it quasiparticles}.

A priori, a slave boson $\phiAn$ can be introduced for any pair $(A,n)$.
However, in this paper, we shall restrict ourselves to phases which
do not display an off-diagonal superconducting long-range order, and hence
one can restrict the $\phiAn$'s to pairs of states which have the same total
particle number on a given site (the local hamiltonian $\Hloc$ commutes
with $\sum_\a \dd_\a d_\a$). The formalism is easily extended to superconducting
states~\cite{fresard_rotinv_intjmodphys_1992,attaccalite_gutzwiller_prb_2003,
bulka_negUslave_prb_1996} by lifting this assumption and modifying appropriately 
the expressions derived in this section. In the following, we consider basis states
$A$ which are eigenstates of the local particle number
(denoted by $N_A$), and hence a $\phiAn$ is introduced provided
$\sum_\a n_\a$$=$$N_A$.

The representation of such a basis state in the enlarged Hilbert space is
defined as:
\beq
|\underline{A}\ket \equiv \frac{1}{\sqrt{D_A}}
\sum_n \phi^\dagger_{An}\vacu \otimes \ketn_f\quad.
\label{eq:def_state}
\eeq
In this expression, $D_A$ denotes the dimension of the subspace of the
Hilbert space with particle number identical to that of $A$, i.e., 
$D_A$$\equiv D(N_A)$$=$$\binom{M}{N_A}$. This insures a proper normalization of 
the state. As before, the ``underline'' in $|\underline{A}\ket$ allows to 
distinguish this state, which lives in the tensor product Hilbert space of 
QP and boson states, from the physical electron state $\ketA$.

Having decided on the physical states, we need to identify a set of
constraints which select these physical states out of the enlarged
Hilbert space in a necessary and sufficient manner. It turns out that the 
following ($M^2+1$) constraints achieve this goal:
\begin{eqnarray}
\sum_{An}\phi^\dagger_{An}\phiAn&=&1
\label{eq:constraint_one}\\
\sum_A \sum_{nn'} \phi^\dagger_{An'}\phiAn\,
\bran \fd_\a f_{\a'}^{\hfill}|n'\ket&=&f_\a^{\dagger}\,f_{\a'}^{\hfill}\quad,
\forall\,\a\quad.
\label{eq:constraint_two}
\end{eqnarray}

The first constraint is obvious and requires that the physical states are
single-boson states. It is easy to check that the physical states satisfy the
second set of constraints (\ref{eq:constraint_two}), but a little more subtle
to actually prove that this set of constraints is sufficient to uniquely select
the physical states (\ref{eq:def_state}) in the enlarged Hilbert space.
The detailed proof is given in Appendix~\ref{appx:constraints}.
Let us emphasize that the order of primed and unprimed indices
in (\ref{eq:constraint_two}) is of central importance.

\subsection{Representation of the physical electron operators}
\label{sec;physical_operators}

We now turn to the representation of the physical electron creation operator on the
representatives (\ref{eq:def_state}) of the physical states in the enlarged 
Hilbert space. We need to find an operator which acts on these representatives
exactly as $\dd_\a$ acts on the physical basis $\ketA$. Namely, given the matrix
elements $\braA\dd_\a\ketB$ such that
\beq
\dd_\a\,\ketB = \sum_A\, \braA\dd_\a\ketB\, \ketA\quad,
\eeq
we want to find an operator $\underline{d}^\dagger_\a$ (in terms of the boson and
QP operators) such that
\beq
\underline{d}^\dagger_\a\,|\underline{B}\ket = \sum_A\, \braA\dd_\a\ketB\,
|\underline{A}\ket\quad.
\label{eq:action_physical_op}
\eeq

\subsubsection{Proximate expression}
\label{sec:simple_expression}

As in the case of the density-density interactions discussed above 
(Sec.~\ref{sec:KR_bosons}), the answer is not unique. We first construct the 
generalization of expression (\ref{eq:creation_op_KR}) to the present formalism
(i.e., ignore at first the question of the proper operators to be inserted
in order to recover the correct non-interacting limit). The following expression is
shown in Appendix~\ref{appx:physical} to satisfy (\ref{eq:action_physical_op}):
\beq
\underline{d}^\dagger_\a=
\sum_{\b,AB,nm}\frac{\braA\dd_\a\ketB\bran\fd_\b\ketm}{\sqrt{N_A(M-N_B)}}
\, \phi^\dagger_{An}\phi_{Bm}^{\hfill}\,\fd_\b\quad.
\label{eq:creation_op}
\eeq
We note that $N_A$$=$$N_B+1$ in this expression can take the values 
$1,\ldots,M$.

Hence, we see that within this formalism, the physical and QP operators are
indeed related by a non-diagonal transformation (\ref{eq:defR}):
\beq
\ud_{\,\a}\,=\, \hat{R}[\phi]_{\a\b}\,f_\b
\eeq
with the $\hat{R}$-matrix corresponding to (\ref{eq:creation_op}) given by
($\hat{R}^*_{\a\b}$ denotes the complex conjugate of $\hat{R}_{\a\b}$):
\beq
\hat{R}[\phi]^{*}_{\a\b}=
\sum_{AB,nm}\frac{\braA\dd_\a\ketB\bran\fd_\b\ketm}{\sqrt{N_A(M-N_B)}}
\, \phi^\dagger_{An}\phi_{Bm}^{\hfill}\quad.
\eeq
The action of (\ref{eq:creation_op}) on physical states, and the proof
that it satisfies (\ref{eq:action_physical_op}) are detailed in
Appendix~\ref{appx:physical}.

\subsubsection{Improved expression}
\label{sec:improved_expression}

The simple expression (\ref{eq:creation_op}), although having the correct action
on the physical states, suffers from the same drawback than 
(\ref{eq:creation_op_KR}) in the case of density-density interactions. Namely, at 
saddle-point level (i.e., with the constraint satisfied on average instead of 
exactly), the non-interacting limit is not appropriately recovered. Thus, one 
needs to generalize the improved expression (\ref{eq:creation_op_KR_norm}) to the 
present rotationally-invariant formalism. However, care must be taken to do so in 
a way which respects gauge invariance (i.e., the possibility of making an 
arbitrary unitary rotation on the QP orbital indices, see 
Sec.~\ref{sec:gauge_invariance}).

We consider the following operators, bilinear in the bosonic fields:
\begin{eqnarray}
\DD^{(p)}_{\a\b} &\equiv& \sum_{Anm}\phi^\dagger_{An}\phi_{Am} 
\bra m|\fd_\a f_\b^{\hfill}|n\ket \,\\
\DD^{(h)}_{\a\b} &\equiv& \sum_{Anm}\phi^\dagger_{An}\phi_{Am}
\bra m|f_\b^{\hfill} \fd_\a|n\ket \quad,
\end{eqnarray}
which can be interpreted as particle- and hole- like QP density matrices
(note that when the constraint is satisfied exactly:
$\DD^{(h)}_{\a\b}$$=$$\delta_{\a\b}-\DD^{(p)}_{\a\b}$).
We then choose to modify the $R$-matrix in the following manner (see 
Appendix \ref{appx:physical}):
\begin{eqnarray}
&&\hspace{-0.6cm}\hat{R}[\phi]_{\a\b}^{*}=\hspace{-0.2cm}
\sum_{AB,nm,\gamma} \braA\dd_\a\ketB
\bra n|\fd_\gamma|m\ket\,
\phi^\dagger_{An}\phi_{Bm}^{\hfill}\,M_{\gamma\b}\,,\label{eq:creation_op_norm}\\
&&\hspace{-0.8cm}\mbox{with}\,\,
M_{\gamma\b}\equiv\left\langle\gamma
\left|\left[\frac{1}{2}
\left(\DD^{(p)}\DD^{(h)}+\DD^{(h)}\DD^{(p)}\right)\right]^{-\frac{1}{2}}
\right|\beta\right\rangle\,.
\end{eqnarray}
We chose to let the QP density matrices enter the $M$-matrix in a symmetrized way 
in order to respect equivalent treatment of particles and holes.  Expression 
(\ref{eq:creation_op_norm}) can be shown to be gauge-invariant, and turns out to
yield the correct non-interacting limit at saddle point.
However, although it yields a saddle point satisfying all
the appropriate physical requirements, it is not fully justified as
an operator identity.

\subsection{Gauge invariance}
\label{sec:gauge_invariance}

As usual in formalisms using slave particles, a {\it gauge symmetry}
is present which allows one to freely rotate the QP orbital indices, 
independently on each lattice site. Physical observables are of course
gauge-invariant. Let us consider an arbitrary $SU(M)$ rotation
of the QP operators:
\beq
\fd_\a\,=\,\sum_\b U_{\a\b}^{\hfill}\,\ftd_{\b}
\label{eq:gauge_QP}
\eeq
This rotation induces a corresponding unitary transformation
${\cal U}(U)$ of the QP Fock states $|n\ket_{\rm f}$. This unitary 
transformation is characterized by the fact that the expectation value of 
$f_\a$ in its Fock basis is an invariant tensor: it is the same in every basis. 
Therefore (summation over repeated indices is implicit everywhere in the 
following):
\begin{align}
\label{eq:Invariant_Op1}
\bra n | f^\dagger_\beta |m\ket &= U_{\beta\beta'} {\cal U}(U)^*_{nn'}
\bra n' | f^\dagger_{\beta'} | m' \ket\,{\cal U}(U)_{mm'}\\
\label{eq:Invariant_Op2}
 \moy{n | f^\dagger_\alpha f_\beta | m} &=  U_{\alpha\alpha'} U_{\beta\beta'}^*
  {\cal U}^*(U)_{nn'}   
\moy{n' | f^\dagger_{\alpha'}f_{\beta'}^{\hfill} | m' }{\cal U}(U)_{mm'}
\end{align}
(the second expression can be deduced from the first using closure relations).
We can now check that if the slave bosons transforms like
\begin{equation}
\phi_{An} = {\cal U}(U)_{nn'}\widetilde\phi_{An'}
\label{eq:gauge_bosons}
\end{equation}
then the constraints and the expressions of the physical electron
operator (either (\ref{eq:creation_op}) or (\ref{eq:creation_op_norm})) are
gauge-invariant. Namely, the $R$-matrix obeys the following transformation law:
\begin{equation}
\hat{R}[\phi]_{\a\b}= \hat{R}[\widetilde{\phi}]_{\a\b'}\,U_{\b\b'}
\end{equation}
and therefore the physical electron operator is invariant:
\begin{equation}
d_\alpha=\hat{R}[\widetilde{\phi}]_{\alpha\beta} \widetilde{f}_{\beta} =
\hat{R}[\phi]_{\alpha\beta} f_{\beta}
\end{equation}

\subsection{Change of physical and quasiparticle basis sets}
\label{sec:basis_change}

It is clear that the basis $|A\ket$ of the local Hilbert space 
(i.e., the physical basis states) can be chosen arbitrarily in this formalism. 
Indeed, making a basis change from $|A\ket$ to $\kettA$, all the expressions 
above keep an identical form provided the bosons corresponding to the new basis 
are defined as:
\beq
\phi^\dagger_{\widetilde{A}n}\,=\,\sum_A \bra A|\widetilde{A}\ket\,
\phi^\dagger_{An}
\label{eq:basis_change}
\eeq
As mentioned above, it is often convenient to use  the eigenstates $|\G\ket$ 
of $\Hloc$ as a basis set.

Changing the basis states associated with {\it quasiparticles} is a somewhat 
trickier issue. Up to now, we have worked with Fock states $|n\ket_f$. A 
different basis set $|Q\ket_f$ can be used, provided however the unitary 
matrix $\bra Q|n\ket$ is {\it real}, i.e., $\bra Q|n\ket$$=$$\bra n|Q\ket$. 
Indeed,
the matrix element $\bra Q|n\ket$ appears in the transformation of the physical 
states and of the constraint, while $\bra n|Q\ket$ appears in the transformation 
of the physical electron operator. When this matrix elements are real, new bosons 
can be defined in the transformed QP basis according to:
\beq
\phi^\dagger_{AQ}\,=\,\sum_n \bra Q|n\ket\,
\phi^\dagger_{An}\,\,\,,\,\,\,(\bra Q|n\ket =\bra n|Q\ket)
\label{eq:basis_change_QP}
\eeq
In particular, when the local hamiltonian is a real symmetric matrix, the same
linear combinations of Fock states which define the atomic multiplets $|\G\ket$ can
be used for QPs, and bosons $\phi_{\G\G'}$ can be considered. This is 
sometimes a useful way of interpreting the formalism and the results at
saddle point (see Sec.~\ref{sec:two_orbitals}).

\subsection{Expression of the hamiltonian, free energy and Green's function}
\label{sec:ham_and_free}

In this section, we derive the expression of the hamiltonian in terms of the 
slave boson and QP fermionic variables. We then construct the free-energy 
functional to be minimized within a mean-field treatment, and express the 
Green's function and self-energy at saddle point.

We recall that the full hamiltonian (\ref{eq:ham_generic}) reads, in terms of the 
physical electron variables: $H$$=$$H_{\rm{kin}}+\sum_i H_{\rm{loc}}[i]$ with
$H_{\rm{kin}}$$=$$\sum_\vk\sum_{\a\b}\varepsilon_{\a\b}(\vk)\,\dd_{\vk\a} d_{\vk\b}$ the 
intersite kinetic energy and $H_{\rm{loc}}$ the local part of the hamiltonian on 
a given site $i$, with general form (\ref{eq:hloc_generic}).

It is easily checked that the following bosonic operator is a faithful 
representation of $H_{\rm{loc}}$ on the representatives of the physical states in 
the enlarged Hilbert space:
\beq
\underline{H}_{\,\rm{loc}}\,=\,
\sum_{AB} \braA H_{\rm{loc}} \ketB\, \sum_n\phi^\dagger_{An}\phiBn
\label{eq:rep_hamloc}
\eeq
If the basis $|\G\ket$ of atomic multiplets is used, this simplifies down to:
\beq
\underline{H}_{\,\rm{loc}}\,=\,
\sum_{\G} E_\G\,\sum_n\phi^\dagger_{\G n}\phi_{\G n}^{\hfill}
\label{eq:rep_hamloc_atomic_basis}
\eeq
Using the bosonic $R$-operators relating the physical electron to the QP
operators, yields the following expression of the kinetic energy:
\beq
\underline{H}_{\,\rm{kin}}\,=\,
\sum_\vk \sum_{\a\a'\b\b'} [\hat{R}^\dagger]_{\a\a'}
\varepsilon_{\a'\b'}(\vk)\hat{R}_{\b'\b}\,
\fd_{\vk\a} f_{\vk\b}^{\hfill}
\label{eq:rep_kinetic}
\eeq
A mean-field theory is obtained by condensing the slave bosons into
c-numbers $\langle\phi_{An}\rangle \equiv \varphi_{An}$. The
constraints are implemented by introducing Lagrange multipliers:
$\lambda_0$ associated with (\ref{eq:constraint_one}) and
$\l_{\a\a'}\equiv [\Lambda]_{\a\a'}$
associated with (\ref{eq:constraint_two}). The saddle point is obtained by
extremalizing, over the $\varphi_{An}$'s and the Lagrange multipliers,
the following free-energy functional:
\begin{eqnarray}
&&\Omega[\{\varphi_{An}\};\Lambda,\lambda_0]\,=\,\\ \nonumber
&&=-\frac{1}{\beta} \sum_\vk \rm{tr}
\ln\left[1+e^{-\beta\left(\mathbf{R}^\dagger(\varphi)
\efat(\vk)\mathbf{R}(\varphi)+\Lambda\right)}
\right]-\lambda_0\\ \nonumber
&&\hspace{0.4cm}+\sum_{ABnn'}\varphi^*_{An'}\,
\bigl\{\,\delta_{nn'}\delta_{AB}\,\lambda_0+\delta_{nn'}\braA H_{\rm{loc}} \ketB\\
&&\hspace{0.4cm}-\delta_{AB}\sum_{\alpha\beta} \Lambda_{\alpha\beta}
\bra n|f^\dagger_\alpha f_\beta |n'\ket\,\bigr\}\,
\varphi_{Bn}
\label{eq:gpot}
\end{eqnarray}
The saddle-point equations, as well as technical aspects of their numerical
solution, are detailed in Appendix~\ref{appx:saddle}.

Finally, we derive the expressions of the Green's functions $\hat{G}$, the 
self-energy $\hat{\Sigma}$ and the QP weight $\hat{Z}$ at saddle point. For the 
QPs, the one-particle Green's function 
$G_{f,\a\b}(\vk,\tau-\tau')$$\equiv$$ -\langle\fd_{\vk\a}(\tau)f_{\vk\b}(\tau')\rangle$ reads (in matrix form):
\beq
\mathbf{G}_f^{-1}(\vk,\omega)\,=\,\omega-
\mathbf{R}^\dagger(\varphi)\,\efat(\vk)\,\mathbf{R}(\varphi) -\mathbf{\Lambda}
\label{eq:Green_QP}
\eeq
and hence the physical electron Green's function reads (we drop the $\varphi$
dependence for convenience):
\begin{eqnarray}
\hspace*{-0.6cm}\mathbf{G}_d^{-1}(\vk,\omega)&=&
[\mathbf{R}^\dagger]^{-1} \mathbf{G}_f^{-1}\mathbf{R}^{-1} \nonumber\\
&=&\omega\,(\mathbf{R R^\dagger})^{-1}\,
-[\mathbf{R}^\dagger]^{-1}\mathbf{\Lambda}\mathbf{R}^{-1}
-\efat(\vk)\,\,,
\label{eq:Green_physical}
\end{eqnarray}
while the non-interacting Green's function is (including the one-body
term present in $\Hloc$):
\beq
\mathbf{G}_{d0}^{-1}(\vk,\omega)=
\omega\openone-\efat^0-\efat(\vk)\quad.
\label{eq:Green_free}
\eeq
The physical self-energy is thus:
\begin{eqnarray}
\hspace*{-0.6cm}\mathbf{\Sigma}_d(\omega)&\equiv&
\mathbf{G}_{d0}^{-1}-\mathbf{G}_d^{-1}\nonumber\\
&=&\omega\left(1-[\mathbf{R R^\dagger}]^{-1}\right)\,
+[\mathbf{R}^\dagger]^{-1}\mathbf{\Lambda}\mathbf{R}^{-1}-\efat^0\,\,.
\label{eq:Sigma_physical}
\end{eqnarray}
So that the matrix of QP weights is obtained in terms of the
$\hat{R}$-matrix at saddle point as:
\beq
\mathbf{Z}=\mathbf{R R^\dagger}\quad.
\label{eq:QP_weight}
\eeq
This generalizes (\ref{eq:Z_simple}) to non-diagonal cases.
It is easily checked that these expressions of the physical quantities
$\mathbf{G}_d,\mathbf{\Sigma}_d$ and $\mathbf{Z}$ are indeed gauge-invariant.

\section{Illustrative results}
\label{sec:results}

In the following, we apply the above formalism to three different
model problems in strongly correlated physics. First, we consider two popular 
models, namely the two-band Hubbard model on a three-dimensional (3D) cubic 
lattice, and a ``bi-layer'' model, coupling two Hubbard 3D cubic lattices . Finally 
a two-site cluster (cluster-DMFT) approximation to the single-band Hubbard model 
on a two-dimensional (2D) square lattice is investigated. 
Hence these models have in common that they all involve two coupled orbitals
(associated, in the cluster-DMFT (CDMFT) framework, to the dimer made of
two lattice sites). The present formalism is of course not restricted to
two-orbital problems, however such models provide the simplest examples where 
the power of the method may be demonstrated.

\subsection{Two-band Hubbard model\label{sec:two_orbitals}}

\begingroup
\begin{table*}[t]
\caption{Eigenstates $|\Gamma\rangle$ of the $SU(2)$ rotationally-invariant 
two-band Hubbard model. Spin values and energies are given for the eigenstates. 
The last column shows the slave bosons for the description of the eigenstates in 
the SBMFT formalism.\label{table-states}}
\begin{ruledtabular}
\begin{tabular}{rccccc}
No. & Eigenstate $|\Gamma\rangle$ & $S_{\Gamma}$ & $S^z_{\Gamma}$ &
$E_\Gamma$ & $\phisb_{\Gamma n}$\\ \hline\hline
1  & $|00,00\rangle$ & 0 & 0 & 0 & $\phisb_{1,|00,00\rangle}$ \\ \hline
2  & $|\spinup 0,00\rangle$ & $\frac{1}{2}$ & $\frac{1}{2}$ & 0 &
     $\phisb_{2,|\spinup 0,00\rangle}$ \\
3  & $|0\spindown,00\rangle$ & $\frac{1}{2}$ & -$\frac{1}{2}$ & 0 &
     $\phisb_{3,|0\spindown,00\rangle}$ \\
4  & $|00,\spinup 0\rangle$ & $\frac{1}{2}$ & $\frac{1}{2}$ & 0 &
     $\phisb_{4,|00,\spinup 0\rangle}$ \\
5  & $|00,0\spindown\rangle$ & $\frac{1}{2}$ & -$\frac{1}{2}$ & 0 &
     $\phisb_{5,|00,0\spindown\rangle}$ \\[0.1cm] \hline
6  & $|\spinup 0,\spinup0\rangle$ & 1 & 1 & $U'-J$ &
     $\phisb_{6,|\spinup 0,\spinup 0\rangle}$ \\
7  & $\frac{1}{\sqrt{2}}\left(|\spinup 0,0\spindown\rangle+|0\spindown,\spinup 0\rangle\right)$ & 1 & 0 & $U'-J$ &
$\left(\phisb_{7,|\spinup 0,0\spindown\rangle},\,\phisb_{7,|0\spindown,\spinup 0\rangle}\right)$  \\
8  & $|0\spindown,0\spindown\rangle$ & 1 & -1 & $U'-J$ &
     $\phisb_{8,|0\spindown,0\spindown\rangle}$ \\
9  & $\frac{1}{\sqrt{2}}\left(|\spinup 0,0\spindown\rangle-|0\spindown,\spinup 0\rangle\right)$ & 0 & 0 & $U'+J$ &$\left(\phisb_{9,|\spinup 0,0\spindown\rangle},\,\phisb_{9,|0\spindown,\spinup 0\rangle}\right)$ \\
10 & $\frac{1}{\sqrt{2}}\left(|\spinup\spindown,00\rangle-|00,\spinup\spindown\rangle\right)$ & 0 & 0 & $U-J_C$ & $\left(\phisb_{10,|\spinup\spindown,00\rangle},\,\phisb_{10,|00,\spinup\spindown\rangle}\right)$ \\
11 & $\frac{1}{\sqrt{2}}\left(|\spinup\spindown,00\rangle+|00,\spinup\spindown\rangle\right)$ & 0 & 0 & $U+J_C$ & $\left(\phisb_{11,|\spinup\spindown,00\rangle},\,\phisb_{11,|00,\spinup\spindown\rangle}\right)$ \\[0.1cm] \hline
12 & $|\spinup\spindown,\spinup 0\rangle$ & $\frac{1}{2}$ & $\frac{1}{2}$ &
     $U+2U'-J$ &
$\phisb_{12,|\spinup\spindown,\spinup 0\rangle}$ \\
13 & $|\spinup\spindown,0\spindown\rangle$  & $\frac{1}{2}$ & -$\frac{1}{2}$ &
     $U+2U'-J$ &
$\phisb_{13,|\spinup\spindown,0\spindown\rangle}$ \\
14 & $|\spinup 0,\spinup\spindown\rangle$ & $\frac{1}{2}$ & $\frac{1}{2}$ &
     $U+2U'-J$ &
$\phisb_{14,|\spinup 0,\spinup\spindown\rangle}$ \\
15 & $|0\spindown,\spinup\spindown\rangle$ & $\frac{1}{2}$ & -$\frac{1}{2}$ &
     $U+2U'-J$ &
$\phisb_{15,|0\spindown,\spinup\spindown\rangle}$ \\[0.1cm] \hline
16 & $|\spinup\spindown,\spinup\spindown\rangle$ & 0 & 0 & $2U+4U'-2J$ &
     $\phisb_{16,|\spinup\spindown,\spinup\spindown\rangle}$ \\
\end{tabular}
\end{ruledtabular}
\end{table*}
\endgroup

The Hubbard model involving two correlated bands, without further onsite
hybridization or crystal-field splitting, serves as one of the standard
problems in condensed matter theory. In contrast to the traditional single-band
model, the formal interaction term in Eq.~(\ref{eq:hloc_generic}) now generates
in the most general fully $SU(2)$ symmetric case four energy parameters, i.e.,
the intraorbital Hubbard $U$ and the interorbital Hubbard $U'$ as well as the
two exchange couplings $J$,$J_C$. Thus the present atomic hamiltonian reads
\begin{eqnarray}
H_{\rm loc}&=&U\sum_{\a} n_{\a\uparrow}n_{\a\downarrow}+U'\sum_{\sigma\sigma'}
n_{1\sigma}n_{2\sigma'}\nonumber\\
&&-J\sum_\sigma n_{1\sigma}n_{2\sigma}\nonumber
+J\sum_\sigma d_{1\sigma}^\dagger d_{2\bar{\sigma}}^\dagger d_{1\bar{\sigma}}
d_{2\sigma}\\
&&+J_C\left( d_{1\uparrow}^\dagger d_{1\downarrow}^\dagger
d_{2\downarrow}d_{2\uparrow}+d_{2\uparrow}^\dagger d_{2\downarrow}^\dagger
d_{1\downarrow}d_{1\uparrow}\right)\,.
\label{tband_aham}
\end{eqnarray}
The kinetic energy shall contain only intraband terms for a
basic tight-binding (TB) model for $s$-bands on a 3D simple
cubic lattice with lattice constant $a$. Thus the corresponding
hamiltonian is written as
\beq
H_{\rm{kin}}=-\frac{1}{3}\sum_\s\sum_{\a=1,2}t_\a\,\sum_{i,j}\,
d^\dagger_{i\a\s}d^{\hfill}_{j\a\s}\quad,
\label{hkin-2band}
\eeq
with the eigenvalues
\begin{equation}
\varepsilon_{\a}(\vk)=-\frac{2}{3}\,t_\a\sum_{\mu=xyz}\cos(k_{\mu}a)\quad,
\label{3d-disp}
\end{equation}
where $t_{\alpha}$ denotes the hopping parameter for orbital $\alpha$=1,2. For
convenience, we set $a$=1. The factor 1/3 in eq. (\ref{hkin-2band}) is to 
normalize the total bandwidth
to $W_\a$=$4t_\a$. Because of the cubic symmetry, $U'$=$U-2J$ may be used,
and furthermore we set $J$=$J_C$. This model is similar to the one considered
by B\"{u}nemann {\sl et al.}~\cite{bunemann_gutzwiller_prb_1998}
using a generalized Gutzwiller approximation (see also
[\onlinecite{attaccalite_gutzwiller_prb_2003}]). Our simpler TB description 
exhibits in principle perfect nesting, however this issue is not relevant at the
present level. The 3D two-band Hubbard model is studied to make contact with 
the named previous work and in order to establish the connection between
the Gutzwiller and slave-boson points of view.

\begin{figure}[t]
\includegraphics*[width=8cm]{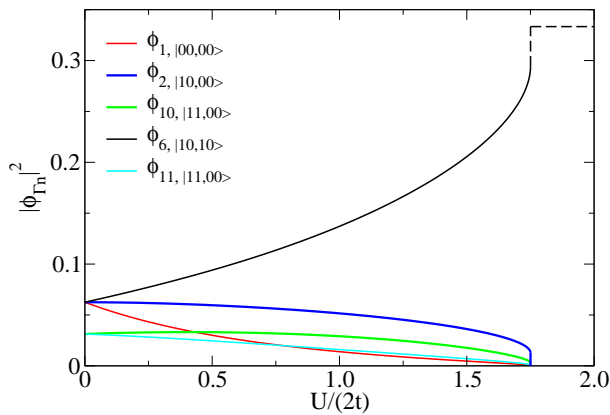}
\caption{(color online)
Inequivalent Slave-boson probabilities $|\phisb_{\Gamma n}|^2$ for the
two-band Hubbard model at half-filling for equal bandwidth and $J/U$=0.2. Note
that $\phisb_{10,|\spinup\spindown,00\rangle}$ and
$\phisb_{11,|\spinup\spindown,00\rangle}$ describe part of the singlet states,
hence their overall amplitude is scaled by 1/$\sqrt{2}$.
\label{hfprob1.fig}}
\end{figure}

\begin{figure}[b]
\includegraphics*[width=8cm]{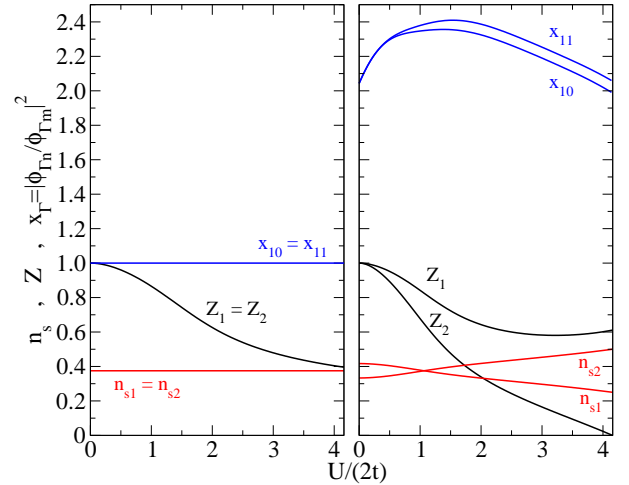}
\caption{(color online)
Comparison of the two-band model for $W_1$=$W_2$ (left) and $W_1$=$2W_2$
(right) at noninteger filling $n$=1.5 and $J/U$=0.2. The ratio $x$ is plotted
for the singlet states coupling the doubly occupied Fock states
(see Tab~\ref{table-states}),
demonstrating that $\varphi_{\G\G'}$ is no longer diagonal in this case.
\label{12dop.fig}}
\end{figure}

When working in the $SU(2)$ rotationally-invariant case,
the $2^4$=16 atomic eigenstates $|\Gamma\rangle$ of the local hamiltonian
(\ref{tband_aham}) serve as the appropriate atomic basis (see
Table~\ref{table-states}), however also the simpler Fock basis (or any other)
may be used. Of course, in the Fock basis, a more complicated energy
{\sl matrix} must be used in the saddle-point equations
(see Appendix~\ref{appx:saddle}). It should be clear from
Tab.~\ref{table-states} that there are 20 nonzero slave-boson amplitudes
$\phisb_{\Gamma n}$ for the current problem. The $S^z$=0 triplet as
well as the three singlets are described with two $\phisb_{\Gamma n}$,
respectively. In principle, even more $\phisb_{\Gamma n}$ may be introduced
in the beginning of the iteration cycle to minimize $\Omega$, but at
convergence those one will come out to be strictly zero. Of course, in 
high-symmetry situations there is still some redundancy within the set of the
20 SBs. For instance, for equal bandwidth at half filling 
(see Fig.~\ref{hfprob1.fig}), all the one- and
three-particle SBs are identical, as well as the zero- and four particle
SB. Moreover the $S^z=\pm$1 triplet SBs are equal because of the
degeneracy. The two SBs describing the $S^z$=0 triplet
are also identical, with a magnitude
$\phisb_{(t,0)n}$=$\phisb_{(t,\pm 1)n}/\sqrt{2}$. Also the bosons
describing one specific singlet have the same absolute value, however they
carry the multiplet phase information, i.e., have plus or minus sign.
In conclusion, in the orbitally degenerate case, the SB amplitudes at
saddle point are of the form:
\begin{equation}
\varphi_{\Gamma n}=\langle n|\Gamma\rangle\, y_\Gamma\quad,
\label{sbamp}
\end{equation}
in which the matrix element $\bra\Gamma|n\ket$ is entirely determined by
$\Hloc$ and $y_\Gamma$ is a (coupling-dependent) amplitude, depending
only on the eigenstate $\Gamma$. This is more clearly interpreted when atomic
states are also used as basis states for QPs (Sec.~\ref{sec:basis_change}).
Indeed, Eq.~(\ref{sbamp}) means that:
\beq
\varphi_{\G\G'}^{\hfill}=\delta_{\G\G'}\,y_\G
\label{eq:atomdiag}
\eeq
Hence, in this highly symmetric case, the saddle point is indeed
of the diagonal form considered in 
Refs.~[\onlinecite{bunemann_gutzwiller_prb_1998,dai06}].
Once the symmetry is lowered, more SBs become inequivalent
and this relation does not hold anymore: there are off-diagonal
components even when the basis of atomic states is used for both physical
and QP states. In this context, the present formalism becomes essential.
Different bandwidths for each orbital, together
with a finite doping away from half-filling lead for instance to two different
absolute values for the two SBs associated with the singlets formed
by the two doubly-occupied Fock states (as seen at the end of this paragraph in
Fig.\ref{12dop.fig}).

Since no interorbital hybridization is applied in this section, the
$\hat{Z}$-matrix is diagonal. We consider first the simple case of
equal bandwidths $t_1$$=$$t_2$$=$$0.5$ (note that in all our applications, 
$t$ sets the unit of energy), thus $Z_{11}$$=$$Z_{22}$$=$$Z$. 
\begin{figure}[t]
\includegraphics*[width=8cm]{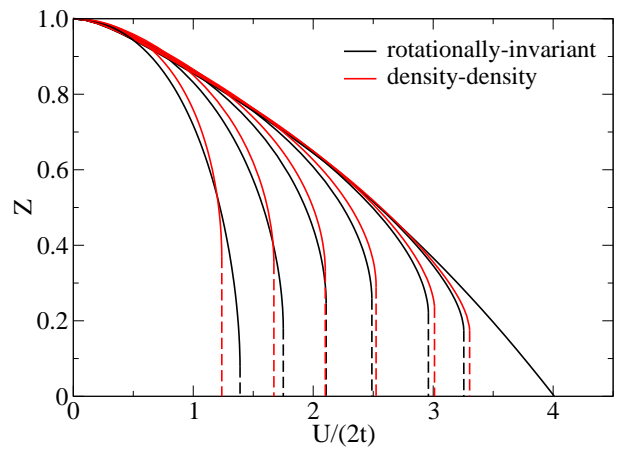}
\caption{(color online)
Influence of $J$ on the Mott transition in the two-band Hubbard model
at half filling ($n$=2) for equal bandwidth. From right to left:
$J/U$=$0,\,0.01,\,0.02,\,0.05,\,0.10,\,0.20,\,0.45$.\label{hfj.fig}}
\end{figure}
\begin{figure}[b]
\includegraphics*[width=8cm]{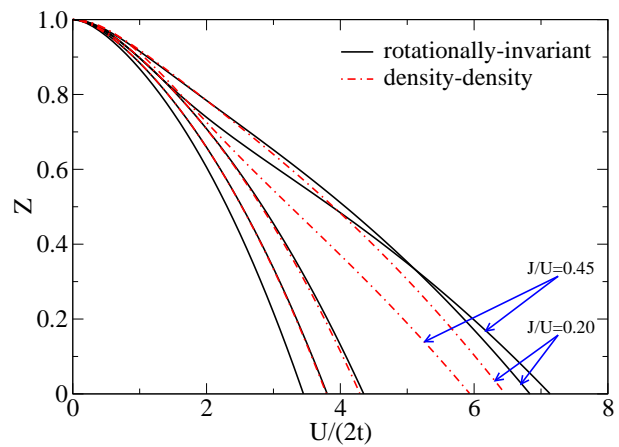}
\caption{(color online)
Influence of $J$ on the Mott transition in the two-band Hubbard model
at quarter filling ($n$=1) for equal bandwidth. The first three combined curves
for the two types of interactions (from left to right) belong to:
$J/U$=$0,\,0.05,\,0.10$. Arrows indicate the labelling for the two larger $J/U$ 
ratios.\label{qfj.fig}}
\end{figure}
Figure~\ref{hfj.fig} shows the variation of $Z$ for different ratios $J/U$ in the 
half-filled case ($n$=2).
The critical coupling $U_c$ for the Mott transition with $J$=0 obtained from
this slave-boson calculation is in accordance with the result of the analytical
formula given by Fr\'esard and Kotliar~\cite{fresard_multiorbital_prb_1997}. It is
seen that an increased $J$ lowers the critical $U$ and moreover changes the
transition from second to first order. Note that in this regard,
Fig.~\ref{hfj.fig} depicts $Z$ up to the spinodal boundary, i.e., the true
transition (following from an energy comparison) is expected to be at slightly 
lower $U_c$. One can also observe the nonmonotonic character for the evolution of 
the critical $Z$ at this boundary when increasing $J/U$. We plot in 
Fig.~\ref{hfj.fig} additionally the results when restricting the atomic 
hamiltonian to density-density terms only, in order
to check for the importance of the then neglected spin-flip and pair-hopping
terms. For larger $J/U$ the critical $Z$ from the latter description is larger
compared to the rotationally-invariant one and moreover it is monotonically 
growing. The latter feature strengthens the first-order character in the 
density-density formulation for growing $J/U$, whereas for rotationally-invariant
interactions this character is strongly weakened in that regime. Although for 
$J/U$=0.45 the jump of $Z$ is quite small, the transition is however still first 
order in the present calculation. Furthermore, there appears to be a crossover 
between the two approaches concerning the reachable metallic spinodal boundary 
when increasing $J/U$.

\begin{figure}[t]
\includegraphics*[width=8cm]{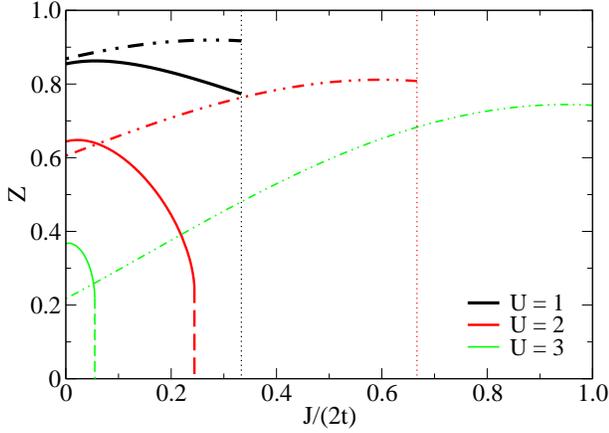}
\caption{(color online)
Influence of $J$ for fixed $U$ at $n$=2 (solid lines) and $n$=1
(dotted-dashed lines) for equal bandwidth and full $SU(2)$ symmetry. The
vertical dotted lines mark the limit we set for $J$, respectively.
\label{j12.fig}}
\end{figure}

At quarter filling ($n$=1) a continuous transition is obtained for all the
previous interactions (see Fig.~\ref{qfj.fig}). Compared to the half-filled case,
the density-density approximation appears to be less severe for small $J/U$, but 
leads to some differences compared to the rotationally-invariant form for large 
$J/U$. Note that for $J/U$=0.45, $U$$'-$$J$ in the local hamiltonian 
(\ref{tband_aham}) becomes negative. Thus a corresponding change of the ground 
state may lead to the resulting nonmonotonic behavior for $U_c$ then observed in 
Fig.~\ref{qfj.fig}. The critical $U$ for $J$=0 is smaller than at half filling and 
with increasing $J/U$ the transition is shifted to {\sl larger} $U_c$ (with the 
above named exception for $J/U$ large). Hence $J$ has a rather different influence 
on the 
degree of correlation for the two fillings. While for $n$=2 the Hund's rule 
coupling substantially enhances the correlations, seen by the decrease in $Z$, 
for $n$=1 the opposite effect may be observed. This is also demonstrated in 
Fig.~\ref{j12.fig} which displays the influence of $J$ for fixed values of $U$ 
comparing half filling with quarter filling.
\begin{figure}[b]
\includegraphics*[width=8cm]{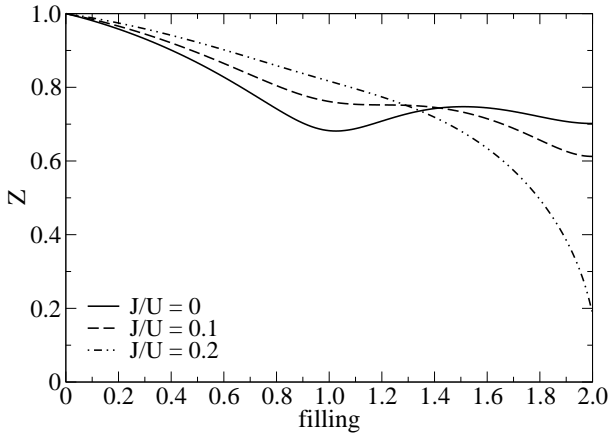}
\caption{Filling dependence of $Z$ for selected values of $J/U$ within the
equal-bandwidth two-band model with full $SU(2)$ symmetry ($U$=1.75).
\label{jdop.fig}}
\end{figure}
The strong decrease in $Z$ upon increasing $J$ was recently shown to
be important for the physical properties of actinides, in particular regarding
the distinct properties of $\delta$-Plutonium and Curium~\cite{shi07}.
For each $U$ shown in Fig.~\ref{j12.fig}, the density-density limiting value
$U/3$ was used as an upper bound for $J$. However, for $U$=2 and $U$=3 the
system shows already a first-order transition at half-filling below the latter
limit.

The QP residue $Z$ is shown as a function of filling $n$ in Fig.~\ref{jdop.fig}
for $U$=1.75 and three ratios $J/U$. For $J$=0 it is observed that $Z(n)$
exhibits two minima, both located at integer filling. The minimum at $n$=1 is 
deeper, corresponding to a lower value for $U_c$ in the quarter-filled case. 
Because of the filling-dependent effect of $J$ seen in Fig.~\ref{j12.fig}, the 
nonmonotonic character of $Z(n)$ is lifted for growing $J/U$.

The two-band Hubbard model was already extensively studied in the more elaborate 
DMFT framework in infinite dimensions. Such
investigations reveal the same qualitative change of the critical $U$ for 
different integer fillings~\cite{roz97,ono03}, of course with some minor 
quantitative differences. Also the 
reduction~\cite{han98,ono03,pru05,son05,ina06} of $U_c$ and the onset of a 
first-order Mott transition~\cite{ono03,pru05,son05,ina06} for finite $J/U$ at 
half-filling is in accordance. Concerning the latter effect, the trend
of weakening the first-order tendency for large $J/U$ is also reproduced and
there is some discussion~\cite{pru05,ina06} about the possibility of even changing
back to a continuous Mott transition in that regime. The increasing $U_c$ with
growing $J$ at quarter filling was also found by Song and Zou~\cite{son05}.

Finally, in Fig.~\ref{12dop.fig} a comparison between the equal-bandwidth and
the different-bandwidth cases at noninteger filling $n$=1.5 is displayed
($J/U$=0.2). For $W_1$=$W_2$ the model does not show a metal-insulator transition
because of the doping. Also the filling of both bands is identical and constant
with increasing $U$ ($n_{s1}$=$n_{s2}$=0.375), and as stated earlier the SBs
are still of the form given by eq.~(\ref{sbamp}). However, when breaking the
symmetry between the two bands by considering different bandwidths,
the model behaves qualitatively rather differently. The individual band fillings 
are not identical anymore, favoring the larger-bandwidth band for $U$=0. With 
increasing $U$ the system manages to drive at least one band insulating by 
transferring charge from the broader into the narrower band, until the latter 
is filled with one electron~\cite{kog04,rue05}. Hence $Z_2$ of the narrower band 
becomes zero at an orbital-selective Mott transition 
(OSMT)~\cite{ani02,kog04,lie04,demedici_osmt_2005,rue05,fer05}. This asymmetric 
model has also a more sophisticated SB description, since for instance the SBs 
of the singlets built out of the respective doubly-occupied Fock states have now 
different amplitudes.

\subsection{The Hubbard bilayer}
\label{sec:bilayer}
\begin{figure}[b]
\includegraphics*[width=8cm]{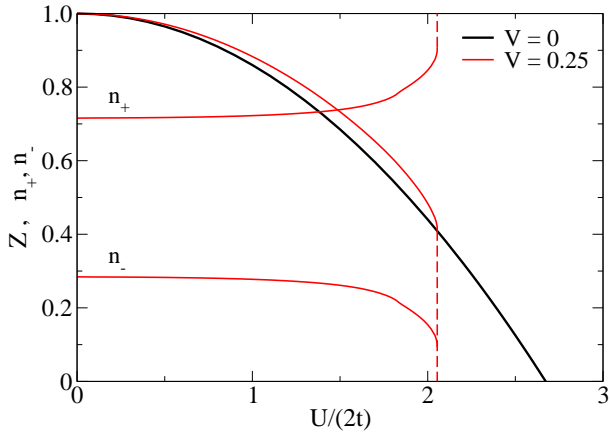}
\caption{(color online)
Half-filled bi-layer with equal bandwidth for $V$=0 and $V$=0.25. For $V$=0 the
filling per spin within the two bands is identical ($n_{s1}$=$n_{s2}$=0.5), whereas
for finite $V$ the the symmetry-adapted bonding/antibonding states have different
filling, denoted $n_+$,$n_-$.}
\label{fig:dl-hf}
\end{figure}

Next, we consider a model consisting of two single-band Hubbard models 
(two ``layers''), coupled by an interlayer hopping $V$. This rather popular model 
has already been subject of various studies~\cite{mon97,moe99,fuh06,kan07}. For 
simplicity  and in order to make connection to the previous section, each layer is 
described here by a 3D cubic lattice, with an onsite repulsion $U$ and an 
intralayer bandwidth $W_\a$ ($\a=1,2$), possibly different for the two layers. 
Hence the local hamiltonian for this problem reads
\begin{eqnarray}\nonumber
H_{\rm loc}=&&U\sum_{\a=1,2} n_{\a\uparrow}n_{\a\downarrow}
+ V \sum_\sigma \left(d_{1\sigma}^\dagger d_{2\sigma}^{\hfill}+
                      d_{2\sigma}^\dagger d_{1\sigma}^{\hfill}\right)\\
&&+\frac{J}{2}\sum_{\sigma\sigma`}
d_{1\sigma}^\dagger d_{1\sigma'}^{\hfill}
d_{2\sigma'}^\dagger d_{2\sigma}^{\hfill}\quad,
\label{eq:dl-ham}
\end{eqnarray}
where the last term describes a possible spin-spin interaction between the
layers. However, for simplicity, we only present in this article results
with $J$$=$$0$. Our choice of kinetic energy is equivalent to the one in the last 
section, i.e., given by Eqs.~(\ref{hkin-2band},\ref{3d-disp}).

In the presence of $V$, an off-diagonal self-energy $\Sigma_{12}(\omega)$
is generated. Furthermore, away from half-filling ($n_1$$+$$n_2$=2), this
self-energy is expected to have a term linear in $\omega$ at low frequency, and
hence $Z_{12}$$\neq$0. We note that, when the bandwidths are equal
($W_1$=$W_2$), the bilayer model can be transformed into a two-orbital model by
a $\vk$-independent rotation to the bonding-antibonding (or $+$,$-$) basis. In
the latter basis, there is no hybridization but instead a crystal-field
splitting (=$2V$) between the two orbitals. The couplings of the
two-orbital hamiltonian are given by (in the notation of the previous section,
and for $J$=0): $U_{\rm eff}$=$U'_{\rm eff}$=$J_{\rm eff}$=$U/2$. When the
bandwidths are different however, the interlayer hopping cannot be eliminated
without generating non-local interdimer interactions.

Due to the reduced symmetry of the present model in comparison to the 
two-band Hubbard model from the previous section, the number of nonzero 
SBs $\phiAn$ equals now 36 (we use here the Fock basis for $|A\rangle$). 
We first consider the simplest case of a half-filled system ($n_1$=$n_2$=1)
with equal bandwidths $W_1$=$W_2$ (and $J$=0). Results for the intralayer
QP weight and the orbital occupancies of the bonding and antibonding bands
are given in Fig.~\ref{fig:dl-hf}. It is seen that the Mott transition is
continuous for $V$=0 but becomes discontinuous in the presence of an
interlayer hopping $V$$\neq$0. For $V$=0.25 the spinodal boundary of the metallic
regime is reached for $U$$\sim$2.055. These results are consistent with findings 
in previous works~\cite{mon97,moe99,fuh06,kan07,cap07} within the DMFT framework.
\begin{figure}[t]
\includegraphics*[width=8cm]{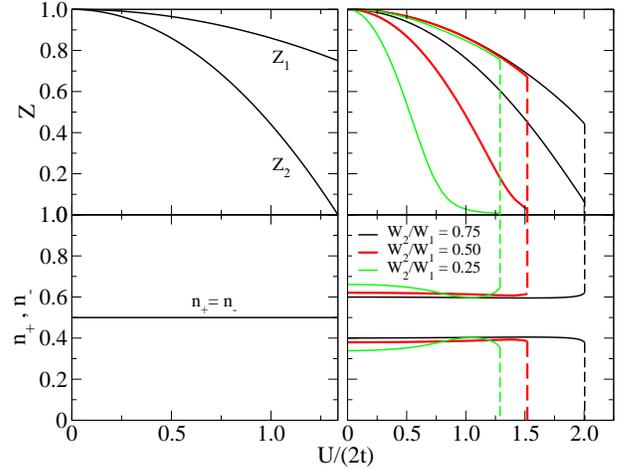}
\caption{(color online) QP residues $Z_i$ and symmetry-adapted fillings 
$n_+$,$n_-$ for the half-filled bilayer. Left: $W_2/W_1$=0.5 and $V$=0.
Right: various bandwidth ratios and $V$=0.1. In the right part, the curves
for smaller $Z$ and $n$ are associated with the lower-bandwidth band.}
\label{fig:dl-unequal-hf}
\end{figure}

Still focusing on the half-filled case, we display in
Fig.~\ref{fig:dl-unequal-hf} the QP weight as a function of $U$ for different
bandwidth ratios $W_2/W_1$. When $V$$=$$0$, one has two independent Mott
transitions in each layer, i.e., an OSMT scenario, at which $Z$ vanishes
continuously. In the presence of a non-zero $V$, this is replaced by a
{\it single discontinuous transition} for both orbitals. This is consistent
with previous findings on the OSMT problem~\cite{demedici_osmt_2005}.
\begin{figure}[t]
\includegraphics*[width=8cm]{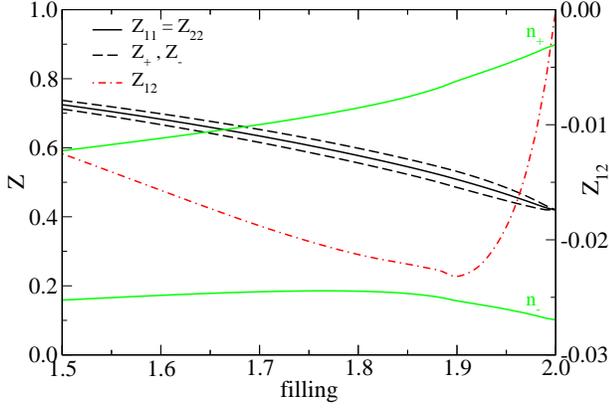}
\caption{(color online) Doped bilayer with equal bandwidth and $V$=0.25 for
$U$=2.054 ($<$$U_c$).}
\label{fig:dl-doping}
\end{figure}
\begin{figure}[t]
\includegraphics*[width=8cm]{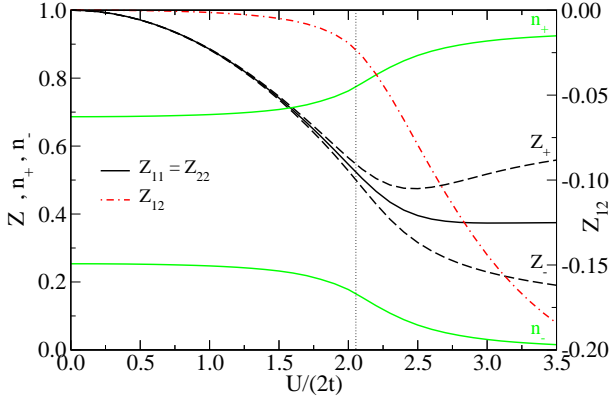}
\caption{(color online) Bilayer at fixed doping ($n$=1.88) with equal bandwidth 
and $V$=0.25. Full green (gray) lines: fillings $n_+$,$n_-$, dashed
dark lines: QP weights $Z_+$, $Z_-$. The vertical dotted line marks
the critical $U$ at half filling.}
\label{fig:dl-uscan}
\end{figure}

We now consider the effect of finite doping away from half filling.
Fig.~\ref{fig:dl-doping} displays the diagonal ($Z_{11}$=$Z_{22}$) elements
as well as the now appearing $Z_{12}$ element of the QP weight matrix as a 
function of doping for $U$$<$$U_c$. Additionally shown are the 
symmetry-adapted QP weights $Z_{+,-}$ (occupations $n_{+,-}$) which follow from 
diagonalizing the $\hat{Z}$ ($\DD^{(p)}$) matrix. Since $Z_{12}$ is small in this
case, the $Z_{+,-}$ are rather similar to $Z_{11}$=$Z_{22}$ and merge with the
latter at half-filling. On the other hand, the polarization of the 
$(+,-)$ bands is still increasing.

\begin{figure}[t]
\includegraphics*[width=8cm]{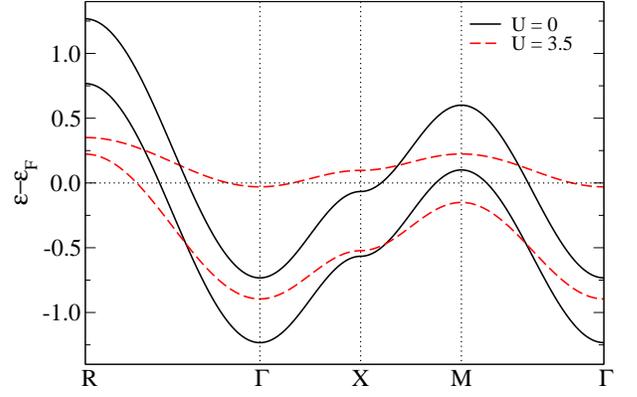}
\caption{(color online) QP bands of the doped bilayer model ($n$=1.88) 
with equal bandwidths and $V$=0.25. The dominately filled band is the bonding
one, respectively.}
\label{fig:dl-bands}
\end{figure}

As it is seen in Fig.~\ref{fig:dl-uscan} the off-diagonal component $Z_{12}$ 
becomes increasingly important for larger $U$ ($>$$U_c$) in the doped case. It 
follows that in this regime the QP weights $Z_{+,-}$ for the 
bonding/antibonding bands have rather different magnitude/behavior. Whereas $Z_-$ 
is monotonically decreasing, $Z_+$ turns around and grows again (as also is the 
filling of the bonding band). Hence, this model is a simple example in which a 
differentiation between QP properties in different regions of the FS occur.
Fig.~\ref{fig:dl-bands} shows the QP ($+$,$-$)-bands in
the noninteracting and interacting case ($U$$>$$U_c$), exhibiting strong
orbital polarization and different band narrowing close to the insulating
state. For very small doping and large $U$ a transition to a new metallic phase is 
found, which will be discussed in detail in a forthcoming publication~\cite{fer07}.
\begin{figure}[b]
\includegraphics*[width=8cm]{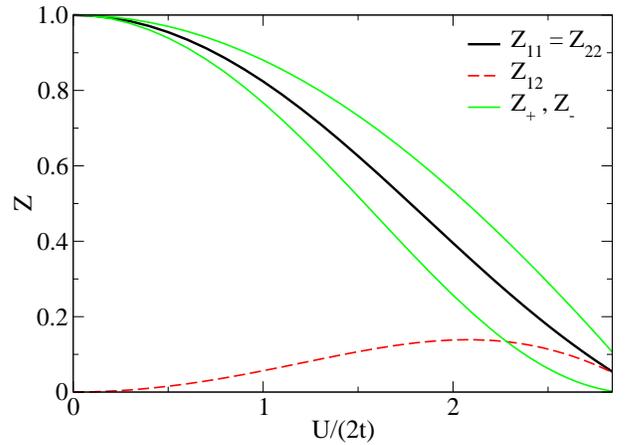}
\caption{(color online)
Half-filled bilayer, with equal bandwidths and $V$=0, but with
a non-local interlayer hybridization $t_{12}$.}
\label{fig:inter-dl}
\end{figure}

Finally, we have also investigated a case in which the interlayer (interorbital)
hopping does not have a local component ($V$=0), but does have a non-local one
$V$=$t_{12}$$\neq$0, treated in the band term of the hamiltonian. Hence the
corresponding energy matrix reads here
\beq
\mbox{\boldmath$\varepsilon$}(\vk) \,=\,
-\frac{2}{3}\,\left(
\begin{array}{cc}
t_{11} & t_{12} \\
t_{12} & t_{22} \\
\end{array} \right)\sum_{\mu=xyz}\cos(k_\mu a)\quad,
\label{eq:Hkin_dl-inter}
\eeq
with the choice $t_{11}$=$t_{22}$=0.5 and $t_{12}$=0.25, as well as $a$=1. 
In that case, a continuous Mott transition within an OSMT scenario can be 
recovered, with, interestingly, a sizeable value of the off-diagonal $Z_{12}$
(Fig.~\ref{fig:inter-dl}). At the transition
$Z_{11}$=$Z_{22}$=$Z_{12}$$\equiv$$Z_c$ holds, i.e., the $\hat{Z}$-matrix has
a zero eigenvalue, associated with the (antibonding) insulating band. Note however
that no net orbital polarization appears with $V$ being purely non-local.

\subsection{Application to the momentum-dependence of the
quasiparticle weight within cluster extensions of DMFT}
\label{sec:cdmft}

In this section, we finally consider the implications of the rotationally
invariant SB technique for the Mott transition and the momentum-dependence of
the QP weight, in the framework of cluster extensions of DMFT.

\begin{figure}[b]
\includegraphics*[width=8cm]{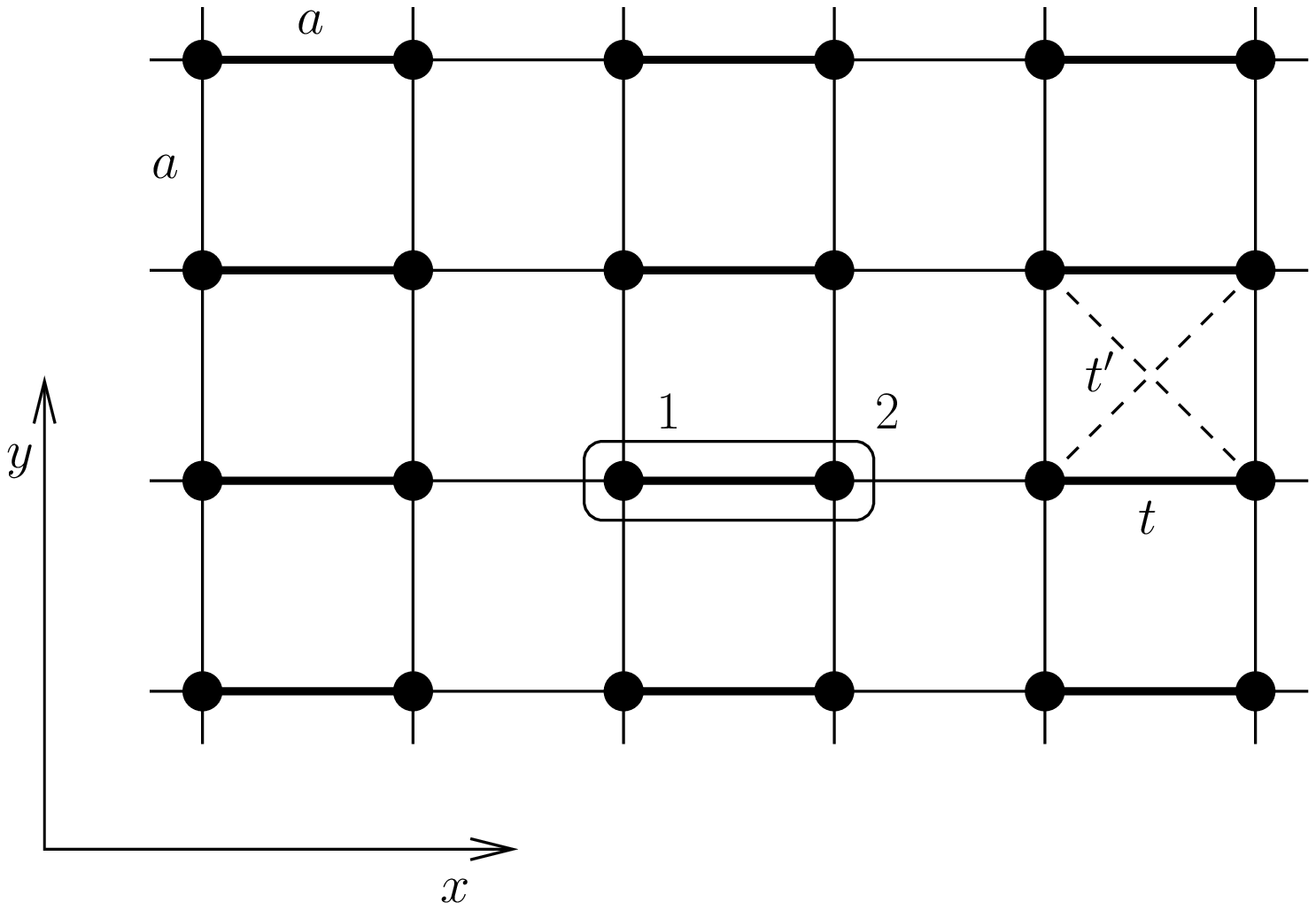}
\caption{Square lattice in the 2-site CDMFT picture.}
\label{fig:tcluster-pic}
\end{figure}
For simplicity, we consider a CDMFT approach to the two-dimensional Hubbard
model with nearest-neighbor hopping $t$ and a next-nearest neighbor hopping $t'$,
based on clusters consisting of two sites (dimers), arranged in a
columnar way on the square lattice (see Fig.~\ref{fig:tcluster-pic}). The
``local'' hamiltonian on each dimer is formally identical to the one
introduced in the previous section for the bilayer model, i.e.
Eq.~(\ref{eq:dl-ham}), with the value $V$=$-t$  of the inter-`orbital' 
hybridization.
The interdimer kinetic energy matrix reads (we set again $a$=1):
\begin{eqnarray}
\varepsilon_{11}(\vK)&=&\varepsilon_{22}(\vK)=-2t\,\cos K_y\\
\varepsilon_{12}(\vK)&=&\varepsilon_{21}^*(\vK)=-t\,\mbox{e}^{i\,2K_x}
-2\tp\,\left(1+\mbox{e}^{i\,2K_x}\right)\cos K_y
\nonumber
\label{eq:Hkin_cdmft_dimer_tp}
\end{eqnarray}
in which $\vK$ denotes a momentum in the reduced Brillouin zone (BZ) of the 
superlattice:
$K_x\in [-\pi/2,+\pi/2]$, $K_y\in [-\pi,+\pi]$. Note again that in SB 
calculations, the intradimer $t$ has to be treated separately
from the rest of the kinetic energy within $H_{\rm loc}$. 
It is easy to check that when putting back $-t$ into the offdiagonal elements of 
the above kinetic-energy matrix, the eigenvalues just correspond to the one of
a single band:
\beq
\varepsilon(\vk)=-2t(\cos k_x +\cos k_y)-4\tp\cos k_x \cos k_y
\label{eq:dis-tprime}
\end{equation}
in the full BZ of the original lattice $k_{x,y}\in [-\pi,+\pi]$.
%
\begin{figure}[t]
\includegraphics*[width=8cm]{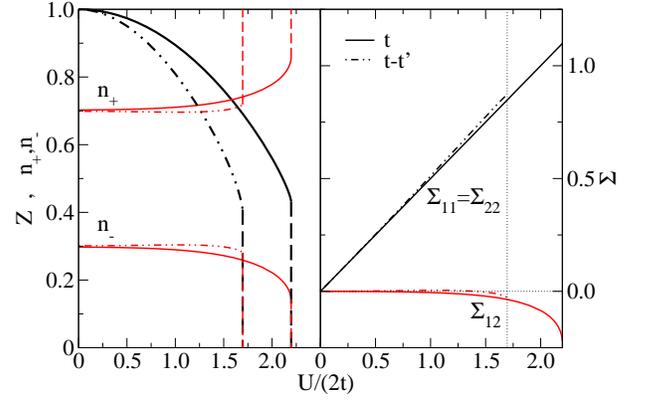}
\caption{(color online) Half-filled two-dimensional Hubbard model
within 2-site CDMFT. Left: QP weights and band fillings, right: static dimer
self-energy $\Sigma$.}
\label{fig:tcluster-hf}
\end{figure}
\begin{figure}[b]
\includegraphics*[width=8cm]{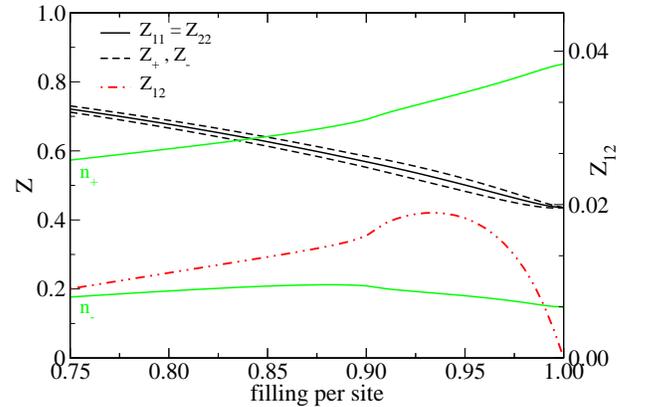}
\caption{(color online) Doped two-dimensional Hubbard model ($\tp$$=$$0$) 
within 2-site CDMFT, for $U$=2.195 ($U_c$$\sim$2.197).}
\label{fig:tcluster-doping}
\end{figure}
%
In the following, we set $t$$=$$0.25$ and consider successively $\tp$$=$$0$ 
and $\tp$$=-0.3\,t$ (a value appropriate to hole-doped cuprates).
Note that, in this article, we do not consider a bigger cluster than the 2-site
dimer, even in the presence of $\tp$. Hence, the cluster self-energy will
only contain $\Sigma_{11}$ and $\Sigma_{12}$ components, i.e., has a spatial
range limited to the dimer. As a result, no renormalization of the
effective $\tp$ is taken into account. This is of course an oversimplification
(particularly in view of the demonstrated physical 
importance~\cite{par04,civelli_breakup_prl_2005} of $\Sigma_{13}$ close to the 
Mott 
transition). Larger clusters will be considered within the present SBMFT in a 
further publication. The goal of the present (simplified) study is to make a 
point of principle, namely that the SB formalism can indeed produce
a momentum-dependent $Z(\vk)$. As the cluster symmetry of the problem at hand
is identical to the bilayer model from the last section, the number of nonzero SBs 
amounts again to 36. 
%
\begin{figure}[t]
\includegraphics*[width=8cm]{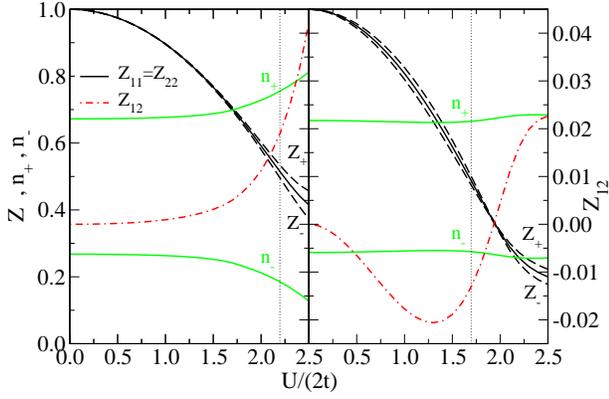}
\caption{(color online) Two-dimensional Hubbard model within 2-site CDMFT at
fixed doping $n$=0.94. Left: $t$-only model, right: $t$-$\tp$ model 
($\tp$$=$$-0.3t$). The vertical lines denote the critical $U$ at half filling, 
respectively.}
\label{fig:ttcluster-uscan}
\end{figure}
Figures~\ref{fig:tcluster-hf}, \ref{fig:tcluster-doping} and
\ref{fig:ttcluster-uscan} summarize the main findings, at half-filling and as
a function of doping, respectively.
Let us first concentrate on the case $\tp$$=$$0$. Obviously, the Mott transition 
at half-filling in this case occurs in a manner which is very similar to the 
bilayer model with a finite interlayer hybridization studied in the previous 
section (Fig.~\ref{fig:dl-hf}): a {\it first-order} transition is found.
The static part of the self-energy $\Sigma_{11}$=$\Sigma_{22}$ equals $U$/2,
while $\Sigma_{12}$ (which has no frequency dependence at half-filling within
SBMFT) has a more complicated negative amplitude close to the transition. Note that
we only discuss the paramagnetic solution, though, of course, the system 
is in principle unstable against antiferromagnetic order for any $U$. Upon doping,
a finite value of $Z_{12}$ is generated. The behavior of $Z_{12}$ is rather 
similar to the case of the bilayer model, except for the change of sign 
(Figs.~\ref{fig:tcluster-doping}, ~\ref{fig:ttcluster-uscan}). Hence
its amplitude is again significantly enhanced for $U$$>$$U_c$, i.e., the $Z_{+,-}$
values tend to manifestly deviate from each other. This is therefore
signalling an increasingly nonlocal component of $Z(\vk)$ as the Mott insulating 
state is approached at strong coupling. Again, further studies in the latter 
regime at small doping will be published soon~\cite{fer07}.  

Including the effect of a nonzero nearest-neighbor hopping $\tp$$\neq$$0$ turns 
out to lead to significant differences.
Although the first-order character of the transition remains stable, the critical 
$U$ is significantly lower (Fig.~\ref{fig:tcluster-hf}). The static components of 
the self-energy behave rather similarly to the $t$-only case, with some minor 
quantitative differences. There is a small negative $Z_{12}$ with a maximum 
amplitude $\sim$0.02, remaining nonzero also at the Mott transition ($\sim$0.01). 
The main difference in comparison to $\tp$$=$$0$ is that here, in the doped case, 
$Z_{12}$ changes sign from negative to positive close to the insulating regime for 
$U$$>$$U_c+\delta$ (with $\delta$$>$$0$) (see Fig.~\ref{fig:ttcluster-uscan}).
Thus the degree of correlation of the effective (bonding-antibonding) bands
is inverted. These differences have to be 
interpreted with caution however, since again the hopping range on the lattice is 
larger than our cluster size, and definitive conclusions will have to be drawn from
a study involving $\Sigma_{13}$ as well.
\begin{figure}[b]
\includegraphics*[width=7cm]{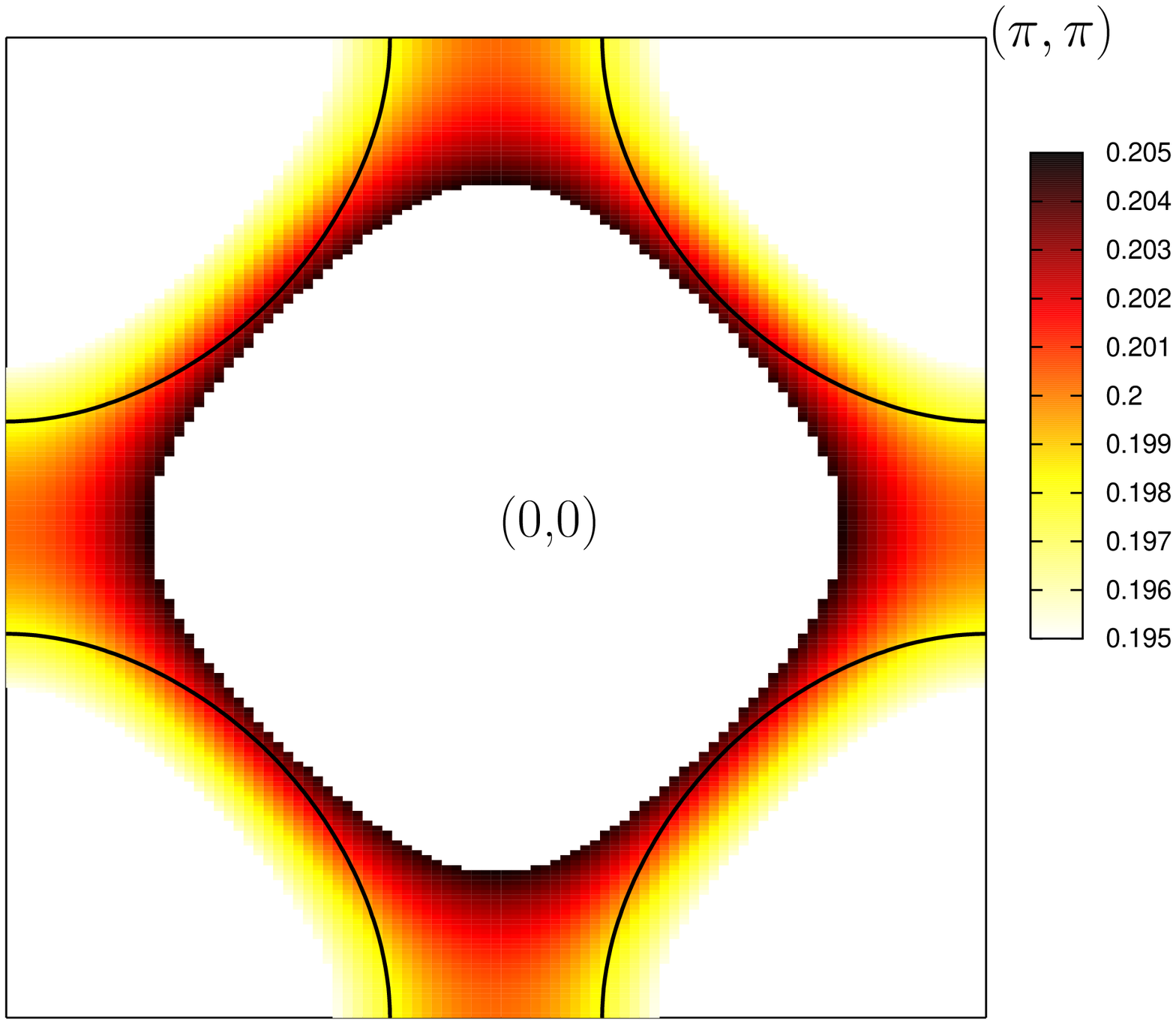}
\caption{(color online) Interacting Fermi surface (solid lines) for the CDMFT
treatment of the 2D $t$-$t'$ Hubbard model with $\tp$$=$$-0.3t$ and $U$=2.5 at
$n$$=$$0.94$ (per site). The color contours show the variation of $Z(\vk)$
(smallest at antinodes).}
\label{fig:zkplot}
\end{figure}

Nonetheless, keeping with the simplified treatment based on a 2-site cluster,
we now describe the resulting momentum dependence of the QP weight
$Z(\vk)$ for the $t$-$\tp$ model. The matrix elements $\Sigma_{11}$ and
$\Sigma_{12}$ of the cluster (physical) self-energy matrix $\mathbf{\Sigma}_c$ are
obtained from the SB amplitudes at saddle point according to 
(\ref{eq:Sigma_physical}).
The self-energy is then periodized on the whole lattice,
in the form~\cite{kot01_2,biroli_clusterlong_prb_2004}:
\begin{equation}
\Sigma_{\rm lat}(\vk,\omega)=\Sigma_{11}(\omega)+\frac{1}{2}\,\Sigma_{12}(\omega)
(\cos k_x +\cos k_y)\,.
\label{eq:siglat}
\end{equation}
The interacting FS is defined as follows
\begin{equation}
\mu-\varepsilon(\vk)-\Sigma_{\rm lat}(\vk,\mbox{$\omega$=0})=0\quad.
\end{equation}
For our case, using Eqs.~(\ref{eq:dis-tprime},~\ref{eq:siglat}), this
reads
\begin{eqnarray}
&&\mu-\Sigma_{11}(0)+
\left[2t-\frac{1}{2}\,\Sigma_{12}(0)\right](\cos k_x +\cos k_y)\nonumber\\
&&\hspace{2cm}+4\tp\cos k_x \cos k_y=0\,\,.
\end{eqnarray}
Hence the FS deforms in a nontrivial way in the presence of $\Sigma_{\rm lat}$,
when including $t'$ in the present 2-site CDMFT description.
The QP weight
$Z(\vk)$ can be derived from $\Sigma_{\rm lat}$ according to:
\begin{equation}
Z(\vk)=\left.\left[1-\frac{\partial}{\partial\omega}
\Sigma_{\rm lat}(\vk,\omega)\right]^{-1}\right|_{\vk=\vk_F}\quad,
\end{equation}
which leads here to:
\begin{eqnarray}
Z(\vk)&=&
\left[[\mathbf{Z_c}^{-1}]_{11}+\frac{1}{2}[\mathbf{Z_c}^{-1}]_{12}
(\cos k_x +\cos k_y)\right]^{-1}\\ \nonumber
&=&\left(Z_{11}^2-Z_{12}^2\right)
\left[Z_{11}-\frac{1}{2}Z_{12}(\cos k_x +\cos k_y)\right]^{-1}\,\,.
\end{eqnarray}
A contour plot of this function is displayed in Fig.~\ref{fig:zkplot}.
Note that it varies only according to $(\cos k_x +\cos k_y)$.
Because the interacting FS involves both $t$ and $\tp$, and hence both
lattice harmonics $(\cos k_x +\cos k_y)$ and $\cos k_x \cos k_y$ (for
$\tp$$\neq$$0$), it cuts through different contour lines of $Z(\vk)$. This
results in a QP weight which varies on the FS.
Figure~\ref{fig:zkplot} shows $Z(\vk)$ for $\vk$ close to the interacting FS.
Albeit the momentum variation is quantitatively quite small, the key qualitative
effect of $Z$ being different on different part of the FS is indeed found.
It is seen that the QPs along the nodal direction,
i.e., along (0,0)-$(\pi,\pi)$, have slightly larger $Z$ than the ones in the the 
antinodal direction ((0,0)-$(0,\pi)$). Hence these results are indeed in 
qualitative agreement with ARPES measurements on cuprates. Note that to get nodal 
points to be more coherent that antinodal ones in this 2-site scheme, 
$Z_{12}$$>$$0$ is actually crucial.

Our results provide, to our knowledge, the first example of a SB calculation
which can address the issue of the momentum dependence of the QP weight.
We believe that the too small variation of $Z$ along the FS found here is
due to the oversimplified 2-site description in which $\Sigma_{13}$ is
neglected. We intend to consider improvements on this issue using
the present SBMFT in a forthcoming work.

Finally, let us make contact with previous work on the two-dimensional Hubbard
model. Of course, this model has been intensively studied with a variety of
methods such as: quantum Monte-Carlo~\cite{hir85,fur92}, exact 
diagonalization~\cite{fan90,dag92}, path-integral renormalization 
group~\cite{kas01}, functional renormalization group~\cite{roh05} and various 
quantum cluster methods (dynamical cluster 
approximation~\cite{mai05}, cluster extensions of dynamical mean-field 
theory~\cite{kotliar_review_rmp_2006}, and variational cluster perturbation 
theory~\cite{tremblay_review_jltp_2006}). We shall not attempt here a detailed 
comparison between the 
rotationally-invariant SB method (which anyhow is a mean-field technique tailored 
to address low-energy issues) with the results of these numerical methods over the 
whole phase diagram  (note in particular that we have not yet investigated 
long-range ordered phases, such as antiferromagnetism or superconductivity).
Rather, we would like to point out that some recent numerical studies using the
above methods~\cite{senechal_hotspots_prl_2004,par04,civelli_breakup_prl_2005,roh05} have indeed revealed the emergence of momentum-space
differentiation in the two-dimensional Hubbard model. We hope that the 
rotationally-invariant SB method will help understand qualitatively the low-energy 
physics emerging from these results.

\section{Conclusion and perspectives}

In this paper, we extended and generalized the rotationally invariant
formulation of the slave-boson method~\cite{li_rotinv1_prb_1989,
fresard_rotinv_intjmodphys_1992}.

Our formulation achieves two goals: (i) extending the slave-boson
method in order to accommodate the most general
crystal fields, interactions and multiplet structures and
(ii) the development of a technique which can describe
QP weights and Fermi liquid parameters
which vary along the Fermi surface.

The key aspect of the formalism is to introduce slave boson
fields which form a matrix with entries labeled by a
pair of a physical state and a QP state (within
an arbitrary choice of basis set). As a result, a
density matrix is constructed instead of just a probability
amplitude for each state.

While the first objective (i) could also be achieved by
generalizing appropriately the Gutzwiller
approximation~\cite{attaccalite_gutzwiller_prb_2003,ferrero_thesis_2006},
we find the slave-boson approach to be somewhat more flexible,
in the sense that it is a mean field theory
which can in principle be improved by computing fluctuations
around the saddle point.
Our application to  the two band model seems promising.
While further work is needed to benchmark the accuracy
of the rotationally-invariant slave-boson method against
exact quantum impurity solvers, it is clear that already in
the single site multiorbital DMFT setting, our method has numerous advantages.
It obeys the Luttinger theorem even in the presence of
multiplets, and can accommodate full atomic physics information. Furthermore,
the off-diagonal elements of the matrix of QP weights
can be calculated within this method, while the standard slave-boson
or Gutzwiller approximations (using probability amplitudes
instead of a density matrix) cannot achieve this goal.

Our technique achieves the second objective (ii) via a detour, namely the use of
cluster extensions of dynamical mean-field theory in order to reduce the
lattice to a multisite (molecular) impurity problem, to which we
apply our rotationally-invariant slave-boson method as an impurity solver.
Because the intersite matrix elements of the QP weight can be calculated,
it leads on the lattice to a momentum-dependence of the QP
residue $Z(\vk)$.
We successfully demonstrated this point, in the framework of a 2-site CDMFT
study of the single-band 2D Hubbard model. We did find that the
QP weight at the nodes is somewhat larger than at the antinodes, although the
magnitude of this effect is expected to increase within a more realistic
study involving a larger cluster (e.g., a square plaquette), which is left for
future work. 
A major challenge is the direct extension of our slave-boson approach to the
lattice, without resorting to the cluster-DMFT detour.
In this context, we mention that other slave-boson techniques, which introduce
magnetic correlations through the use of link variables
to decouple the superexchange $J$ term~\cite{kotliar_largeN_leshouches_1995},
can be interpreted in terms of a $\vk$ dependent self-energy.
However, within such schemes, the derivative of the self-energy with respect to 
frequency is momentum {\it independent} (in contrast to the static part),
yielding a $\vk$ independent QP residue.
Hence, our approach goes beyond these methods, at least in conjunction with the
cluster-DMFT approach. We hope that having an economic impurity solver based on SBs
will allow us to study larger cluster sizes than feasible with other methods,
and most importantly help us understanding the low-energy physics emerging from
these cluster dynamical mean-field theories.

Finally, we limited our study to slave bosons which do not mix the particle number.
The extension to full charge-rotational invariance and superconductivity is
possible (see 
Refs.~[\onlinecite{fresard_rotinv_intjmodphys_1992,bulka_negUslave_prb_1996}]
in the single-orbital case), and will be presented in a separate paper.
In this context, the slave boson method will incorporate the $SU(2)$ charge 
symmetry and its extension away from half filling considered by 
Wen and Lee~\cite{wen96} and 
the rotationally-invariant slave-boson formalism can serve as a powerful tool for 
interpreting the low-energy physics emerging from plaquette-CDMFT studies of 
this issue.

\acknowledgements
We are grateful to Pablo Cornaglia and Michel Ferrero for very
useful discussions and remarks. As this work was being completed, we learned of 
a parallel effort by Michele Fabrizio~\cite{fab07}, in the framework of the 
Gutzwiller approximation. In particular, the form of the constraints advocated 
in this work  matches our constraints 
(\ref{eq:constraint_one},\ref{eq:constraint_two}) 
in SB language. A.G. also thanks him for discussions. This work has been 
supported by the ``Chaire Blaise Pascal'' (r\'egion Ile de France and Fondation de
l'Ecole Normale Sup\'erieure), the European Union (under contract
``Psi-k f-electrons'' HPRN-CT-2002-00295), the CNRS and Ecole Polytechnique.
G.K. is supported by the NSF under Grant No. DMR 0528969.
%
%
\appendix
\section{Single-orbital case and connection with previous work}
\label{appx:single_orbital}

Here, we briefly consider the single-orbital case ($M$$=$$2$), which also allows 
to make contact with 
Refs.~[\onlinecite{li_rotinv1_prb_1989,fresard_rotinv_intjmodphys_1992}].
These authors introduced in this case a rotationally-invariant formalism,
with the calculation of response functions associated with the saddle point
as their main motivation.
For $M$$=$$2$, the following local basis set can be considered
(whether or not $\Hloc$ is diagonal in this basis):
\begin{eqnarray}
N=0:&& |0\ket\,\,,\nonumber\\
N=1:&& |\s\ket=\dd_\s|0\ket\\
N=2:&& |D\ket=\dd_\spinup\dd_\spindown|0\ket\quad.\nonumber
\end{eqnarray}
Hence, we introduce the following bosons (not mixing sectors with different
particle numbers, i.e., not considering superconducting states):
\beq
\phi_{00}\equiv \phi_E\quad,\quad
\phi_{\s\s'}\quad,\quad
\phi_{\spinup\spindown}\equiv\phi_D\quad.
\eeq
Up to normalizations, the bosons $p^\dagger_{\s\s'}$ introduced
in Ref.~\onlinecite{li_rotinv1_prb_1989} correspond to $\phi^\dagger_{\s\s'}$.
In contrast, the standard Kotliar-Ruckenstein~\cite{kotliar_4bosons_prl_1986} 
formalism introduces only {\it two} bosons $p^\dagger_\s$ in the one-particle 
sector. The representatives (\ref{eq:def_state}) of the physical states read here:
\begin{eqnarray}
|\underline{0}\ket&=&\phi_E^\dagger\vacu\nonumber\\
|\underline{\s}\ket&=&\frac{1}{\sqrt{2}}\sum_{\s'}
\phi_{\s\s'}^\dagger\fd_{\s'}\vacu\\
|\underline{D}\ket&=&\phi_D^\dagger\dd_\spinup\dd_\spindown\vacu\quad,\nonumber
\end{eqnarray}
and the constraints (\ref{eq:constraint_one},\ref{eq:constraint_two}) read:
\begin{eqnarray}
1&=&\phi_E^\dagger\phi_E^{\hfill}+\sum_{\s\s'}\phi^\dagger_{\s\s'}\phi_{\s\s'}+
\phi^\dagger_D\phi_D^{\hfill}\\
\fd_\a f_\a^{\hfill}&=&\phi_D^\dagger\phi_D^{\hfill}
+\sum_\s\phi^\dagger_{\s\a}\phi_{\s\a}^{\hfill}\\
\fd_{\spinup}f_{\spindown}^{\hfill}&=&
\sum_\s\phi^\dagger_{\s\spindown}\phi_{\s\spinup}^{\hfill}\\
\fd_{\spindown}f_{\spinup}^{\hfill}&=&
\sum_\s\phi^\dagger_{\s\spinup}\phi_{\s\spindown}^{\hfill}\quad.
\end{eqnarray}
Not including, for simplicity, the square-root normalizations in
(\ref{eq:creation_op_norm}), needed however in order to insure a correct 
$U$$=$$0$ limit at saddle point, the `simplest' expression (\ref{eq:creation_op})
of the electron creation operators read:
\begin{eqnarray}
\underline{d}^\dagger_{\spinup}=\frac{1}{\sqrt{2}}\sum_\b\,
[\phi^\dagger_{\spinup\b}\phi_E+(-1)^\b\phi^\dagger_D
\phi_{\spindown\overline{\b}}]\,\fd_\b\\
\underline{d}^\dagger_{\spindown}=\frac{1}{\sqrt{2}}\sum_\b\,
[\phi^\dagger_{\spindown\b}\phi_E-(-1)^\b\phi^\dagger_D
\phi_{\spinup\overline{\b}}]\,\fd_\b\quad.
\end{eqnarray}
Apart from the motivations of Ref.~[\onlinecite{li_rotinv1_prb_1989}] 
(associated with fluctuations and response functions), the usefulness of the 
rotationally-invariant scheme in the single-orbital case can be demonstrated on 
a toy model consisting of a one-band Hubbard model with a magnetic field, 
purposely written in the $S^x$ direction (i.e., in the form
$h\,d^\dagger_\uparrow d_\downarrow +\rm{h.c}$, analogous to a hybridization).
Although the direction of the field should not matter,
a direct application of the standard Kotliar-Ruckenstein formalism is impossible
in that case. The rotationally-invariant formalism can be shown to lead to the 
correct saddle point, independently of the spin-quantization axis.


\section{Derivation of Eq. (\ref{eq:constraint_two})}
\label{appx:constraints}

In this section, we show that the physical states of the form (\ref{eq:def_state})
are exactly those selected by the constraints (\ref{eq:constraint_two}) and  
(\ref{eq:constraint_one}).
First, it is easy to check that states of the form (\ref{eq:def_state}) do 
satisfy these constraints. Indeed, let us act on the state
$|\underline{C}\ket \equiv \frac{1}{\sqrt{D_C}}
\sum_m \phi^\dagger_{Cm}\vacu \otimes \ketm_f$ and
with (\ref{eq:constraint_two}). The l.h.s leads to:
\begin{align}\nonumber
f_\a^{\dagger}\,f_{\a'}|\underline{C}\ket &=
\frac{1}{\sqrt{D_C}} \sum_m \phi^\dagger_{Cm}\vacu \otimes
f_\a^{\dagger}\,f_{\a'}\ketm_f\\
&=\frac{1}{\sqrt{D_C}} \sum_{mm'} \bra m'|f_\a^{\dagger}\,f_{\a'}\ketm\,
\phi^\dagger_{Cm}\vacu \otimes |m'\ket_f\quad.
\end{align}
When acting with the r.h.s, only the term $A$$=$$C$ and $n$$=$$m$ gives a 
non-vanishing contribution, hence:
\begin{eqnarray}
&&\sum_A \sum_{nn'} \phi^\dagger_{An'}\phi_{An}\,
\bran \fd_\a f_{\a'}|n'\ket
|\underline{C}\ket=\nonumber\\
&&\hspace{1.5cm}\frac{1}{\sqrt{D_C}} 
\sum_{nn'} \bran f_\a^{\dagger} f_{\a'}|n'\ket\,
\phi^\dagger_{Cn'}\vacu \otimes \ketn_f\quad.\nonumber
\end{eqnarray}

We now prove that (\ref{eq:constraint_two}) are {\it sufficient} conditions, 
which is a bit more difficult. Since (\ref{eq:constraint_one}) excludes states 
with more than one boson, it is enough to consider a general state of the form:
\beq
|C;W\ket\equiv\,\sum_{pq} W_{pq}\,\phi^\dagger_{Cp}\vacu \otimes |q\ket_f\quad.
\eeq
and to show that (\ref{eq:constraint_two}) implies $W_{pq}\propto\delta_{pq}$. 
Acting on this state with each term in the constraint 
(\ref{eq:constraint_two}) yields for the l.h.s:
\begin{eqnarray}
f_\a^{\dagger}\,f_{\a'}^{\hfill}|C;W\ket &=&
\sum_{pq} W_{pq}\,\phi^\dagger_{Cp}\vacu 
\otimes \fd_\a f_{\a'}^{\hfill} |q\ket_f\nonumber\\
&&\hspace{-2cm}=\sum_{pr} \phi^\dagger_{Cp}\vacu \otimes |r\ket_f \sum_q W_{pq}
\bra r|\fd_\a f_{\a'}^{\hfill} |q\ket\quad.
\label{eq:lhs_constraint}
\end{eqnarray}
Let us now act with the r.h.s. Only the terms with $A$$=$$C$ and $n$$=$$p$ 
contribute, leading to:
\begin{eqnarray}
&&\sum_A \sum_{nn'} \phi^\dagger_{An'}\phi_{An}\,
\bran \fd_\a f_{\a'}|n'\ket
|C;W\ket\nonumber\\
&&\hspace{0.5cm}=\sum_{pqn'} W_{pq}\,\phi^\dagger_{Cn'}\vacu \otimes
|q\ket_f \bra p|\fd_\a f_{\a'} |n'\ket\nonumber\\
&&\hspace{0.5cm}=\sum_{pr} \phi^\dagger_{Cp}\vacu \otimes |r\ket_f \sum_q W_{qr}
\bra q|\fd_\a f_{\a'} |p\ket\,\,,
\end{eqnarray}
where the last expression comes from a change of indices
$n'\rightarrow p, q\rightarrow r, p\rightarrow q$.

We see that the constraint is satisfied provided that the following identity holds,
{\it for all orbital indices} $\a\a'$ {\it and all} states $p,r$:
\beq
\sum_q W_{pq} \bra r|\fd_\a f_{\a'} |q\ket =
\sum_q W_{qr} \bra q|\fd_\a f_{\a'} |p\ket\quad.
\label{eq:condW1}
\eeq
Let us first look at the case $\a$$=$$\a'$, which reads:
\beq
r_\a\,W_{pr}=p_\a\,W_{pr}\quad.
\eeq
Hence $W_{pr}$$=$$0$ unless $p_\a$$=$$r_\a$ for all $\a$, so that:
\beq
W_{pq}=w_p\,\delta_{pq}\quad.
\eeq
Substituting this into (\ref{eq:condW1}), we obtain:
\beq
w_p\,\bra r|\fd_\a f_{\a'} |p\ket =
w_r\, \bra r|\fd_\a f_{\a'} |p\ket\quad.
\eeq
Thus, if $r$ and $p$ are related by a move of a QP 
from one state to another (a transposition of two occupation numbers),
then $w_{p}$$=$$w_{r}$. Moreover, two Fock states in the same sector $H_{N}$
of the Hilbert space are related by a permutation of the occupied states,
which can be decomposed in a product of transpositions. Hence, $w_{p}$
is a constant for $p \in H_{N}$, and $W_{pq}$$\propto$$\delta_{pq}$ as claimed.


\section{Physical creation operator}
\label{appx:physical}
\subsection{Proximate expression}

First let us note that there is a systematic route to find the expression for $d$, 
which consists in writing the operator as : 
\begin{eqnarray}
\underline{d}^\dagger_\a
&=&\sum_{AB} \braA\dd_\a\ketB\, |\underline{A}\ket\bra\underline{B}|\nonumber\\
&=&\sum_{AB}\sum_{n\in H_A,m\in H_B}\hspace{-0.3cm} \braA\dd_\a\ketB\,
\phi^\dagger_{An}\phi_{Bm}\,X^{\rm f}_{nm}\,\,,
\end{eqnarray}
with $X^f_{nm}$$=$$|n\ket_f\bra m|_f$ in usual Hubbard notations.
$X^f_{nm}$ is obviously {\it not} just a one-particle operator 
$f^\dagger$, even 
when restricted to the sectors of interest in the above formula, since the states 
$n$ and $m$ can differ in many places. However, because any transposition of two 
QPs, {\it when acting on a physical state}, can be replaced by a 
corresponding operation on bosons using the constraint (\ref{eq:constraint_two}), 
and because any product of bosonic operators which cannot be reduced to a 
quadratic form will produce a state which is out of the physical subspace, the
physical operator {\it must} in the end take the form:
\beq
\underline{d}^\dagger_\a\,=\,
\sum_\b\sum_{AB}\sum_{nm} C_{Bm}^{An}(\a,\b)\,\,
\phi^\dagger_{An}\phi_{Bm}^{\hfill}\,\fd_\b\quad.
\label{eq:creation_op_appx4}
\eeq
One can solve for the coefficients $C_{Bm}^{An}(\a,\b)$, requesting
proper action on the physical states. 

In this section however, we restrict ourselves to proving that 
Eq. (\ref{eq:creation_op}) does the job, {\it i.e.}, that 
\beq
\underline{d}^\dagger_\a=
\sum_{\b,AB,nm}\frac{\braA\dd_\a\ketB\bran\fd_\b\ketm\
}{\sqrt{N_A(M-N_B)}}\, \phi^\dagger_{An}\phi_{Bm}^{\hfill}\,\fd_\b
\eeq
satisfies
\beq
\underline{d}^\dagger_\a\,|\underline{B}\ket = \sum_A\, \braA\dd_\a\ketB\,
|\underline{A}\ket\quad.
\eeq
We start by proving the formula : 
\begin{equation}
\sum_\b\sum_{p\in H_{N}} \bran \fd_\b|p\ket\, \fd_\b|p\ket = (N+1) \, \ketn
\label{eq:fermionic_lemma}\quad,
\end{equation}
where the sum over $p$ runs over the basis of the subspace $H_{N}$ 
(states with $N$ QPs) of the Fock space. First, we have in general : 
\begin{equation}
\sum_\b\sum_{p\in H_{N}} \bran \fd_\b|p\ket\, \fd_\b|p\ket = 
\sum_{n'\in H_{N+1}}  a_{n'}  |n' \ket \quad,
\end{equation}
but, because $f^{\dagger}$ is the creation operator (it connects one basis state 
to only one another) :  
\begin{equation}
a_{n'} = \sum_{\b,p\in H_{N}} \bran \fd_\b|p\ket 
\bra n' | \fd_\b|p\ket  \propto \delta_{nn'}  
\sum_{\b,p\in H_{N}}  \left | \bran \fd_\b|p\ket \right |^{2}
\end{equation}
We now use the fact that the tensor is invariant (it has the same expression in 
every basis) and use the notations introduced for Eq. (\ref{eq:Invariant_Op1}):  
$U$ is a unitary transformation of the one QP states and  ${\cal U}$ 
is the corresponding transformation in the Fock states.
$\bra n | f^\dagger_\beta |m\ket$$=$$U_{\beta\beta'} {\cal U}^*_{nn'}
\bra n' | f^\dagger_{\beta'} | m' \ket\,{\cal U}_{mm'}$ and we have : 
\begin{eqnarray}
a_{n } &=& \sum_\b\sum_{m\in H_{N}}  \left | \bran \fd_\b|m\ket 
\right |^{2}\nonumber\\
&=&  \sum_{\b,  m\in H_{N}}  
   \sum_{\empile{\b' m'\in H_{N}}{n' \in H_{N+1}}}   
\sum_{\empile{\b'' m''\in H_{N}}{n'' \in H_{N+1}}}
U_{\b\b'} U_{\b\b''}^{*} {\cal U}_{nn'}^{*}
{\cal U}_{nn''}^{}\nonumber\\   
&&\hspace{1.5cm}\times\,{\cal U}_{mm'}^{}   {\cal U}_{mm''}^{*}  
 \bra n'|  \fd_{\b'} |m'\ket \bra n''|  \fd_{\b''}|m''\ket^{*}\nonumber \\[0.2cm]
 &=&\sum_{\beta'}  \sum_{p\in H_{N}}   \sum_{n',n'' \in H_{N+1}}
{\cal U}_{nn'}^{*}   {\cal U}_{nn''}^{} 
 \bra n'|  \fd_{\b'}|p\ket \bra n''|  \fd_{\b'}|p\ket^{*}\nonumber \\
&=&   \sum_{\b, p\in H_{N}}   \left | \bra ({\cal U}n) | \fd_\b|m\ket \right |^{2} 
\end{eqnarray}
Moreover, any couple of elements of the basis of the Fock state can be connected
by a ${\cal U}$ transformation (with a $U$ that permutes the one QP basis state),
therefore $a_{n}$$\equiv$$a$ is a independent of $n$.
$a$ can then be determined by summing (\ref{eq:fermionic_lemma}) over $n$  : 
\begin{eqnarray}
\sum_{n \in H_{N+1}} a_{n} &=& 
\sum_{\beta}\sum_{\empile{ n \in H_{N+1}}{ p\in H_{N}} } 
\left | \bra n | \fd_\b | p \ket \right |^{2} \nonumber\\
 &=& \sum_{ n \in H_{N+1}}  \bra n | \sum_{\b} 
\fd_\b f_{\beta} | n \ket \nonumber\\
 &=&  \sum_{n \in H_{N+1}}  (N+1)\quad,
\end{eqnarray}
leading to $a$$=$$N+1$. This completes the proof of (\ref{eq:fermionic_lemma}).

It is now simple to compute the action of (\ref{eq:creation_op}) : acting on 
$|\underline{C}\ket$$\equiv$$\frac{1}{\sqrt{D_C}}
\sum_p \phi^\dagger_{Cp}\vacu \otimes |p\ket_f$ with this operator,
only the term $B$$=$$C$,$m$$=$$p$ contributes, and we get : 
\begin{eqnarray}
\underline{d}^\dagger_\a |\underline{C}\ket &=&
\frac{1}{\sqrt{D_{C}(N_C+1)(M-N_C)}}\nonumber\\
&&\hspace{-1.5cm}\times\,\sum_{A,n\in H_{N_C+1}}\hspace{-0.25cm}
\braA\dd_\a\ketC\, \phi^\dagger_{An}\vacu\otimes
\sum_\b\sum_{p\in H_{N_C}} \bran \fd_\b|p\ket\, \fd_\b|p\ket\nonumber\\
&&\hspace{-1cm}=\sqrt{\frac{N_{C }+1}{D_{C}(M-N_C)}} \sum_{A,n\in H_{N_C+1}}
\hspace{-0.45cm}\braA\dd_\a\ketC\,\phi^\dagger_{An}\vacu\otimes |n\ket\nonumber\\
& =& \sum_{A} \braA\dd_\a\ketC\, |\underline{C}\ket\quad,
\end{eqnarray}
which is identical to (\ref{eq:action_physical_op}).

\subsection{Improved expression}

\def\Dp{\hat{\Delta}^{(p)}}
\def\Dh{\hat{\Delta}^{(h)}}

In this section we present arguments for the improved formula used in this paper.
First, it is useful to define the ``natural orbitals'' (NO) basis
as the basis which diagonalizes the quasiparticle and quasihole density matrices 
corresponding to the average constraint, which is given by
\GroupeEquations{
\begin{align}
\Dp_{\a\b}[\phi] &\equiv \sum_{Anm}\phi^*_{An}\phi_{Am} \bra m|\fd_\a f_\b|n\ket \\
\Dh_{\a\b}[\phi] &\equiv \sum_{Anm}\phi^*_{An}\phi_{Am} \bra m|f_\b \fd_\a|n\ket \\
&= \sum_{An}\phi^*_{An}\phi_{An} - \Dp_{\a\b}[\phi]\quad. \nonumber
\end{align}
}
Let us denote by $\xi_\lambda,|\lambda\ket$ the eigenvalues and eigenvectors of 
those matrices  : 
\begin{equation}
\Dp_{\a\b} = \sum_\l \xi_\l \bra\a|\l\ket \bra\l|\b\ket\,\,\,,\,\,\,
|\l\ket = \sum_\l \bra\a|\l\ket \, |\a\ket\,\,,
\end{equation}
which is equivalent to use the NO quasiparticle operator  
$\psi_\l^\dagger$ such that :
\begin{equation}
\psi_\l^\dagger \equiv \sum_\a \bra\l|\a\ket \fd_\a\,\,\,,\,\,\,
\bra\psi^\dagger_\l\psi_\mu\ket =\delta_{\l\mu}\xi_\l\quad.
\end{equation}
To be fully explicit, we can consider the particular basis transformation
(in the notations of the section above) $\fd_\a$$=$$U_{\a\l}\psi^\dagger_\l$ 
which rotates to the NOs, and the corresponding rotation on the 
bosons: $\phi_{An}$$=$${\cal U}(U)_{nn'}\Omega_{An'}$. The rotation matrix is:
\begin{equation}
U_{\a\l} = \bra\a|\l\ket\quad,
\end{equation}
and in the NO basis:
\begin{eqnarray}
\sum_{Anm}\Omega_{An}^*\Omega_{Bm} \bra m|\psi^\dagger_\l \psi_\mu|n\ket &=&
\delta_{\l\mu}\sum_{An}\Omega_{An}^*\Omega_{An}n_\l\nonumber\\
&=&\delta_{\l\mu}\xi_\l(\{\Omega_{An}\})\,\,,
\label{eq:dens_NO}
\end{eqnarray}
and:
\begin{equation}
\Dp_{\a\b}[\phi]=\sum_\l U_{\a\l}\,\xi_\l\, [U^\dagger]_{\l\b}\quad.
\end{equation}
The idea is to generalize the Kotliar-Ruckenstein normalization factor in the 
NO basis, where the QP density being diagonal, its probabilistic 
interpretation is more transparent. Hence, the improved expression of $d$ reads: 
\begin{equation}
\underline{d}^\dagger_\a=\hspace{-0.2cm}
\sum_{\l,AB,nm} \frac{\braA\dd_\a\ketB\,\bran\psi^\dagger_\l\ketm}
{\sqrt{\xi_\l(\{\Omega_{An}\})(1-\xi_\l(\{\Omega_{An}\}))}}\,\,
\Omega^\dagger_{An}\Omega_{Bm}\,\psi^\dagger_\l\,.
\label{eq:creation_op_NO}
\end{equation}
Note that the formal square-root normalisation, i.e., $1/\sqrt{N_A(M-N_B)}$,
does not appear in this representation. We can now rotate back to the generic 
basis we started from and use the gauge invariance, 
leading to  :
\begin{eqnarray}
\label{eq:creation_opla2}
\underline{d}^\dagger_\a &=&
\sum_{AB,nm,\beta\gamma}{\cal C}_{Bm}^{An}(\alpha,\beta)\,
\phi^\dagger_{An}\phi_{Bm}^{\hfill}
\sum_\l\frac{\bra\beta|\l\ket\bra\l|\gamma\ket}{\sqrt{\xi_\l(1-\xi_\l)}}
\fd_\gamma\nonumber\\
&&\hspace{-1cm}=\hspace{-0.4cm}\sum_{AB,nm,\beta\gamma}\hspace{-0.3cm} 
{\cal C}_{Bm}^{An}(\alpha,\beta)\,
\phi^\dagger_{An}\phi_{Bm}^{\hfill}\,
\bra\b|[\Dp\Dh]^{-\frac{1}{2}}|\gamma\ket\,\fd_\gamma\,,\\
&&\mbox{with}\quad{\cal C}_{Bm}^{An}(\alpha,\beta)=
\moy{A|d^\dagger_\a|B}\moy{n|f^\dagger_\b|m}\quad.
\end{eqnarray}
Hence this yields the following form for the $R$-matrix:
\begin{equation}
R[\phi]_{\a\b}^{*}=\hspace{-0.3cm}\sum_{AB,nm,\delta}
\hspace{-0.2cm} {\cal C}_{Bm}^{An}(\alpha,\delta)\,
\phi^\dagger_{An}\phi_{Bm}^{\hfill}\,
\bra\delta|[\Dp\Dh]^{-\frac{1}{2}}|\beta\ket\,\,.\\
\end{equation}

In the actual implementation of the saddle point calculations, 
the explicit use of both the quasiparticle and the quasihole density
matrices has been utilized ({\it i.e., not using their relation}). 
Although {\sl at convergence} the different representations yield the same 
values, writing the equations via both, particle and hole density matrix, 
appears to be necessary within the minimization cycle. This is due to the fact 
that the derivatives with respect to the slave bosons have to be symmetric,
 however when using only the particle density matrix (or its eigensystem 
decomposition) for instance the derivative with respect to the empty
boson vanish, although this one exists in the KR case. 
In the end an even more symmetrized form, i.e.,
$\frac{1}{2}(\Dp\Dh+\Dh\Dp)$ was used for the square root in
eq. (\ref{eq:creation_opla2}). Thus defining the following matrix:
\begin{equation}
M_{\gamma\beta}=\left\langle\gamma\left|
\left[\frac{1}{2}({\Dp}{\Dh}+{\Dh}{\Dp})\right]^{-\frac{1}{2}}\right|
\beta\right\rangle\,\,,
\label{mmat}
\end{equation}
the electron operators are written as 
\begin{eqnarray}
\underline{d}^\dagger_\a &=&
\sum_{AB}\sum_{nm}\sum_{\gamma\beta} {\cal C}_{Bm}^{An}(\alpha,\gamma)\,
\phi^\dagger_{An}\phi_{Bm}\,M_{\gamma\beta}\,\fd_\beta\nonumber\\
&=&\sum_\b R_{\a\b}^*\fd_\beta\\
\underline{d}_\a &=&
\sum_{AB}\sum_{nm}\sum_{\gamma\beta} {\cal C}_{Bm}^{An}(\alpha,\gamma)\,
\phi^\dagger_{Bm}\phi_{An}\,M_{\beta\gamma}\,f_\beta\nonumber\\
&=&\sum_\b R_{\a\b}f_\beta\quad,
\end{eqnarray}
and correspondingly independently written the elements of the 
$R$,$R^\dagger$-matrices read
\begin{eqnarray}
&&\hspace{-0.5cm}R[\phi]_{\a\b}=
 \sum_{AB,nm,\gamma}\,{\cal C}_{Bm}^{An}(\alpha,\gamma)\,
\phi^\dagger_{Bm}\phi_{An}^{\hfill}\,\hat{M}_{\beta\gamma}\label{rmatone}\\
&&\hspace{-1.2cm}R^\dagger[\phi]_{\a\b}\equiv R[\phi]_{\b\a}^*=\hspace{-0.3cm}
 \sum_{AB,nm,\gamma}\hspace{-0.3cm}
{\cal C}_{Bm}^{An}(\beta,\gamma)\,\phi^\dagger_{An}\phi_{Bm}^{\hfill}
\,\hat{M}_{\gamma\alpha}\,.
\label{rmattwo}
\end{eqnarray}


\section{Details on the saddle-point equations and their numerical solution}
\label{appx:saddle}

The saddle-point equations for $T$=0 are obtained by performing the partial
derivatives with respect to all the variables, i.e., condensed slave-boson
amplitudes and Lagrange multipliers:
\begin{eqnarray}
\frac{\partial\Omega}{\partial\lambda_0}&=&
1-\sum_{An}\varphi^\dagger_{An}\varphi^{\hfill}_{An}\\
\frac{\partial\Omega}{\partial\Lambda_{\alpha\beta}}&=&
\langle f^\dagger_\a f_\b \rangle-\sum_{A,nn'}\varphi^\dagger_{An'}
\moy{n|d^\dagger_\alpha d_\beta |n'}\,\varphi_{An}\\
\frac{\partial\Omega}{\partial\varphi_{Cm}^{\hfill}}&=&
\sum_{\vk j}\tilde{f}_{\vk j}
\frac{\partial\varepsilon_{\vk j}}{\partial\varphi_{Cm}}+
\sum_{A}E_{AC}^{\hfill}\varphi^\dagger_{Am}+
\lambda_0\varphi^\dagger_{Cm}\nonumber\\
&&-\sum_{\a\b}\Lambda_{\a\b}\sum_{n'}
\moy{m|d^\dagger_\alpha d_\beta |n'}\varphi^\dagger_{Cn'}\\
\frac{\partial\Omega}{\partial\varphi^\dagger_{Cm}}&=&
\sum_{\vk j}\tilde{f}_{\vk j}
\frac{\partial\varepsilon_{\vk j}}{\partial\varphi^\dagger_{Cm}}+
\sum_{B}E_{CB}^{\hfill}\varphi^\dagger_{Bm}+
\lambda_0\varphi_{Cm}^{\hfill}\nonumber\\
&&-\sum_{\a\b}\Lambda_{\a\b}\sum_{n}
\moy{n|d^\dagger_\alpha d_\beta |m}\varphi_{Cn}^{\hfill}\\
&&\hspace{-2cm}\mbox{with}\quad \langle f^\dagger_\a f_\b \rangle=
\sum_{\vk j}\tilde{f}_{\vk j}\moy{\a|\nu_{\vk j}}\moy{\nu_{\vk j}|\b}\quad.
\end{eqnarray}
The $\varepsilon_{\vk j}$ are the eigenvalues (with band index $j$) of the
QP matrix $({\bf R}^\dagger(\varphi)\efat(\vk){\bf R}(\varphi)+
{\bf \Lambda})$ with
corresponding eigenvector $|\nu_{\vk j}\rangle$, while $\tilde{f}_{\vk j}$
denotes the occupation number of the state $|\nu_{\vk j}\rangle$ for a given
total number of particles, to be evaluated by standard $k$-integration
techniques (e.g. tetrahedron method, Gaussian smearing, etc.).

\subsection{Some slave-boson derivatives}
\paragraph{Eigenvalues and $R$ matrices.}
The derivatives of the eigenvalues with respect to the slave bosons, i.e.,
$\frac{\partial\varepsilon_{\vk j}}{\partial\varphi}$ may be performed
pertubatively:
\begin{eqnarray}
\frac{\partial\varepsilon_{\vk j}^{\hfill}}
{\partial\varphi}&=&\moy{\nu_{\vk j}\left|
\frac{\partial}{\partial\varphi}\left({\bf R}^\dagger\efat(\vk){\bf R}+
\Lambda\right)\right|\nu_{\vk j}}\nonumber\\
&=&\moy{\nu_{\vk j}\left|\frac{\partial {\bf R}^\dagger}{\partial\varphi}
\efat(\vk){\bf R}+{\bf R}^\dagger
\efat(\vk)\frac{\partial {\bf R}}{\partial\varphi}\right|\nu_{\vk j}}\nonumber\\
&=&\left\langle\nu_{\vk j}\left|\sum_{\a\b}\ketal
\frac{\partial\hat{R}_{\a\b}^\dagger}{\partial\varphi}\brabe 
\efat(\vk){\bf R}\right.\right.\nonumber\\
&&\left.\left.\hspace{0.7cm}
+\sum_{\a\b}{\bf R}^\dagger\efat(k)|\alpha\rangle\frac{\partial\hat{R}_{\a\b}}
{\partial\varphi}\langle\b\right|\nu_{\vk j}\right\rangle\nonumber\\
&&\hspace{-0.5cm}=\sum_{\a\b}\left[\langle\nu_{\vk j}|\a\rangle
\frac{\partial\hat{R}_{\a\b}^\dagger}{\partial\varphi}
\langle\b|\efat(\vk){\bf R}|\nu_{\vk j}\rangle\right.\nonumber\\
&&\left.\hspace{0.7cm}
+\langle\nu_{\vk j}|{\bf R}^\dagger\efat(\vk)|\a\rangle
\frac{\partial\hat{R}_{\a\b}}{\partial\varphi}
\langle\b|\nu_{\vk j}\rangle\right]\,\,.
\end{eqnarray}
The therefore needed explicit expressions for the derivatives of
the $R$,$R^\dagger$-matrices read as follows
(using eqs.(\ref{rmatone},\ref{rmattwo})):
\begin{eqnarray}
&&\frac{\partial\hat{R}_{\a\b}}{\partial\varphi_{Cm}}=\hspace{-0.3cm}
\sum_{AB,nn',\gamma}\hspace{-0.3cm}
 {\cal C}_{Bn'}^{An}(\a,\gamma)\,\varphi^\dagger_{Bn'}\,
\left[\delta_{An}^{Cm}\hat{M}_{\b\gamma}+\varphi_{An}
\frac{\partial\hat{M}_{\b\gamma}}{\partial\varphi_{Cm}}\right]\nonumber\\
&&\frac{\partial\hat{R}_{\a\b}}{\partial\varphi^\dagger_{Cm}}=\hspace{-0.3cm}
\sum_{AB,nn',\gamma}\hspace{-0.3cm} {\cal C}_{Bn'}^{An}(\a,\gamma)\,\varphi_{An}\,
\left[\delta_{Bn'}^{Cm}\hat{M}_{\b\gamma}+\varphi^\dagger_{Bn'}
\frac{\partial\hat{M}_{\b\gamma}}{\partial\varphi^\dagger_{Cm}}\right]\nonumber\\
&&\left(\mbox{analogous for}\quad [\hat{R}^\dagger]_{\a\b}\right)\nonumber
\end{eqnarray}
\paragraph{The ${\bf M}$ matrix.}
As it is seen, the derivatives involve the derivative of the ${\bf M}$
matrix (\ref{mmat}). This derivative is computed as follows. Lets first write
${\bf M}$ as
\begin{equation}
\hat{M}_{\gamma\beta}=\bra\gamma|{\bf K}^{-1/2}|\beta\ket\quad.
\end{equation}
What we are looking for is the derivative of ${\bf K}^{-1/2}$ with
respect to the SBs. In order to get access to this quantity we use the
identity
\begin{eqnarray}
\hspace{-0.5cm}(\partial_\varphi {\bf K}^{-1/2}){\bf K}^{-1/2}+{\bf K}^{-1/2}
(\partial_\varphi {\bf K}^{-1/2})&=&\partial_\varphi {\bf K}^{-1}
\label{dphik1}\\
\Leftrightarrow\hspace{1cm} {\bf X}{\bf K}^{-1/2}+{\bf K}^{-1/2}{\bf X}&=&
{\bf Y}\quad,
\end{eqnarray}
with ${\bf X}$=$\partial_\varphi {\bf K}^{-1/2}$ and ${\bf Y}$=$\partial_\varphi {\bf K}^{-1}$. We then apply ${\bf P}$ which transforms ${\bf K}$ to its
eigensystem. This yields
\begin{equation}
{\bf X}'\,{\bf L}\,+\,{\bf L}\,{\bf X}'={\bf Y}'\quad,
\end{equation}
where the prime denotes that the quantities defined above are expressed in
that eigensystem, and ${\bf L}$=${\bf P}^\dagger {\bf K}^{-1/2}{\bf P}$.
Since in the eigensystem ${\bf K}^{-1/2}$ is diagonal, i.e., ${\bf L}$ is,
the last equation can be written in components and ${\bf X}'$ determined:
\begin{equation}
X'_{ij}L_j+L_iX'_{ij}=A'_{ij}\quad
\Leftrightarrow\quad X'_{ij}=\frac{A'_{ij}}{L_i+L_j}\,\,.
\end{equation}
Backtransforming to ${\bf X}$=${\bf P}{\bf X}'{\bf P}^\dagger$=$\partial_\varphi {\bf K}^{-1/2}$ yields the
desired derivative of ${\bf K}$ and subsequently of ${\bf M}$. To perform the
described computation we need to know $\partial_\varphi {\bf K}^{-1}$ in
eq. (\ref{dphik1}), however this quantity may be straightforwardly calculated
when starting from the identity ${\bf KK}^{-1/2}$${\bf K}^{-1/2}$=1,
resulting in $\partial_\varphi {\bf K}^{-1}$=$-{\bf K}^{-1}(\partial_\varphi {\bf K}){\bf K}^{-1}$.
\subsection{Mixing}
In order to solve the saddle-point equations, a method to deal with a system
of nonlinear equations ${\bf F}$ as a function of the variables
(slave bosons, Lagrange multipliers) ${\bf x}$ has to be utilized:
\begin{equation}
{\bf F}({\bf x})=0
\end{equation}
In the present work we tested several quasi-Newton techniques (e.g.
Broyden~\cite{bro65}, modified Broyden~\cite{van84}, etc.) to handle this
numerically. Thereby from a starting guess for ${\bf x}$  the variables are
updated via
\begin{equation}
{\bf x}_{\rm (m+1)}={\bf x}_{\rm (m)}+
{\bf J}_{\rm (m)}^{-1}{\bf F}_{\rm (m)}\quad,
\end{equation}
since we want ${\bf F}_{\rm (m+1)}$ to be zero to linear order. The
jacobian ${\bf J}$ is here defined as follows
\begin{equation}
J_{ij}\equiv-\frac{\partial F_i}{\partial x_j}
\end{equation}
and is not calculated exactly (this would involve second derivatives and
would lead to the Newton-Raphson method) but is computed at each step $m$ via
formulae which dictate several constraints on how ${\bf J}$ should evolve.
In our numerical implementation we found the modified Broyden scheme to be well
suited for the so far investigated applications Note that there is usually no need 
for explicitly fixing the gauge for the numerical solution of the saddle-point 
equations. The initial amplitudes of the variational parameters, i.e., slave 
bosons and lagrange multipliers, at the start of the iteration, together with the 
choice of the atomic basis $|A\rangle$, always ensured proper convergence to one 
of the family of solutions within our implementation.

\bibliographystyle{apsrev}
\bibliography{bibag,pubag,bibextra}

\end{document}